\documentclass[aps,twocolumn,superscriptaddress,pra]{revtex4-1}

\usepackage{amsmath}
\usepackage[version=4]{mhchem} 
\usepackage{changes}
\usepackage{soul}
\usepackage{graphicx}
\usepackage[space]{grffile} 
\usepackage{gensymb}
\usepackage{bm}
\usepackage{amssymb,amsfonts,amsmath,amsbsy,bm,t1enc,lipsum,latexsym}
\usepackage[colorlinks=true,citecolor=blue]{hyperref}

\begin{document}

\title{Soliton microcomb based spectral domain optical coherence tomography}

\author{Paul J. Marchand}
\email{paul-james.marchand@polymtl.ca}
\affiliation{Swiss Federal Institute of Technology Lausanne (EPFL), Laboratoire d'optique biom{\'e}dicale (LOB), Lausanne, CH-1015, Switzerland}
\affiliation{Department of Electrical Engineering, {\'E}cole Polytechnique de Montr{\'e}al, Canada}
\author{J. Connor Skehan}
\affiliation{Institute of Physics, Swiss Federal Institute of Technology Lausanne, CH-1015, Switzerland}
\author{Johann Riemensberger}
\affiliation{Institute of Physics, Swiss Federal Institute of Technology Lausanne, CH-1015, Switzerland}
\author{\mbox{Jia-Jung Ho}}
\affiliation{Institute of Physics, Swiss Federal Institute of Technology Lausanne, CH-1015, Switzerland}
\author{Martin H. P. Pfeiffer}
\affiliation{Institute of Physics, Swiss Federal Institute of Technology Lausanne, CH-1015, Switzerland}
\author{Junqiu Liu}
\affiliation{Institute of Physics, Swiss Federal Institute of Technology Lausanne, CH-1015, Switzerland}
\author{Christoph Hauger}
\affiliation{Carl Zeiss Meditec AG, Rudolf-Eber-Stra{\ss}e 11, 73447 Oberkochen, Germany}
\author{Theo Lasser}
\affiliation{Swiss Federal Institute of Technology Lausanne (EPFL), Laboratoire d'optique biom{\'e}dicale (LOB), Lausanne, CH-1015, Switzerland}
\author{Tobias J. Kippenberg}
\email{tobias.kippenberg@epfl.ch}
\affiliation{Institute of Physics, Swiss Federal Institute of Technology Lausanne, CH-1015, Switzerland}

\date{\today}
\maketitle

\noindent\textbf{
Spectral domain optical coherence tomography (SD-OCT) is a widely used and minimally invasive technique for bio-medical imaging \cite{Drexler2015}. SD-OCT typically relies on the use of superluminescent diodes (SLD), which provide a low-noise and broadband optical spectrum. 
Recent advances in photonic chipscale frequency combs \cite{kippenberg2018dissipative, kues2019quantum} based on soliton formation in photonic integrated microresonators provide an chipscale alternative illumination scheme for SD-OCT. Yet to date, the use of such soliton microcombs in OCT has not yet been analyzed.
Here we explore the use of soliton microcombs in spectral domain OCT and show that, by using photonic chipscale \ce{Si3N4} resonators in conjunction with 1300\,nm pump lasers, spectral bandwidths exceeding those of commercial SLDs are possible. We demonstrate that the soliton states in microresonators exhibit a noise floor that is ca. 3 dB lower than for the SLD at identical power, but can exhibit significantly lower noise performance for powers at the milli-Watt level. 
We perform SD-OCT imaging on an \emph{ex vivo} fixed mouse brain tissue using the soliton microcomb, alongside an SLD for comparison, and demonstrate the principle viability of soliton based SD-OCT.
Importantly, we demonstrate that classical amplitude noise of all soliton comb teeth are correlated, i.e. common mode, in contrast to SLD or incoherent microcomb states \cite{Ji:19}, which should, in theory, improve the image quality. Moreover, we demonstrate the potential for circular ranging, i.e.  optical sub-sampling \cite{Siddiqui2012,Siddiqui2018}, due to the high coherence and temporal periodicity of the soliton state. Taken together, our work indicates the promising properties of soliton microcombs for SD-OCT.}

\begin{figure*}[!t]
\includegraphics[width = 0.85\textwidth]{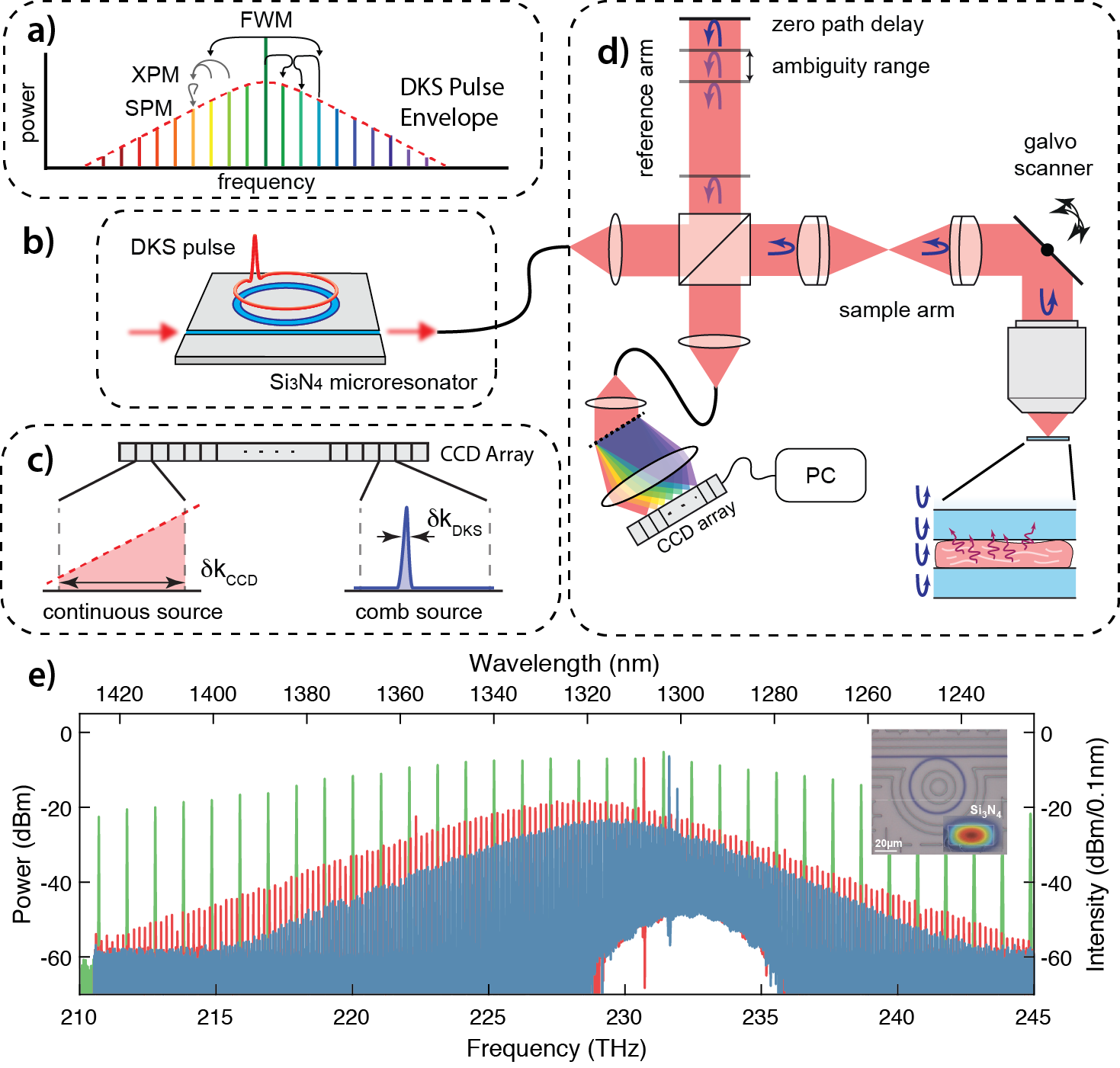}
\caption{\textbf{The principle of dissipative Kerr soliton enabled spectral domain OCT}. a) A dissipative Kerr soliton (DKS), based on the system shown in b), where a CW laser drives nonlinear frequency conversion in a photonic chip-based \ce{Si3N4} micro-resonator. Here, the generated pulse train is comprised of discrete and equally spaced frequency components as determined by the free spectral range of the non-linear cavity. In particular, this process creates a frequency comb via the dual balance between non-linearity and dispersion on one hand, and loss and gain on the other. Eventually, the discrete components of this micro-resonator frequency comb (or continuous source as in traditional OCT) are dispersively projected onto a CCD array as shown in c), after passing through a standard OCT setup as seen in figure d). Experimental data for a variety of free spectral ranges (Green 1 THz, Red 200 GHz, and Blue 100 GHz) typical of micro-resonator DKS are shown in e), along with an inset microscope photograph of a $\sim$1~THz micro-resonator, and an SEM photograph of a typical bus waveguide in \ce{Si3N4}.
}
\label{fig_octPrinciple}
\end{figure*}

First demonstrated in 1991 by Huang \cite{Huang1178}, optical coherence tomography (OCT) has become an important technique for non invasive imaging of biological tissues \cite{Fercher2003}. Today, OCT is a standard diagnostic tool in ophthalmology and has been extended to intravascular imaging \cite{Kassani2017} and brain imaging \cite{Vakoc2009, Bolmont2012a, Srinivasan2012a}. Over the past decade, frequency domain methods (FD-OCT), i.e. spectral-domain OCT (SD-OCT) and swept-source OCT (SS-OCT), have superseded time domain OCT through their higher sensitivity \cite{leitgeb_performance_2003,deBoer:03,Choma:03,Choi,Drexler2004}. Since then, light sources and detectors for FD-OCT (both SD and SS-OCT) have been improved, providing low noise, larger bandwidths and faster acquisition rates. Recently, sources comprised of a set of discrete frequencies have been proposed for FD-OCT, as they offer an increased depth-sensitivity \cite{Tsai2009,Bajraszewski2008a}, reduced power exposure while maintaining a high axial resolution \cite{Jung:08} and an extended imaging range through optical-domain subsampling \cite{Siddiqui2018}. The periodicity in the tomogram offered by this novel acquisition scheme enables significantly extending the OCT imaging range in a data efficient manner and shows great promise for imaging of non-planar samples, such as in intra-operative scenarios \cite{Siddiqui2012}.

One promising implementation of such discrete sources for SD-OCT are soliton microcombs. 
First discovered in 2007, these microcombs are generated by the nonlinear conversion processes inside micro-resonators \cite{DelHaye2007, Kippenberg2011}. Through adjustment of laser power and detuning, a dissipative Kerr soliton (DKS) state can be excited, providing coherence lengths and bandwidths comparable to continuous-wave and femtosecond lasers, respectively \cite{Herr2013a}. The spectrum of a DKS state consists of fully coherent laser lines with linewidths equal to the CW pump laser linewidth (typically $\sim 100\,\mathrm{kHz}$), resulting in kilometer scale coherence lengths. Its overall spectral bandwidth can be tailored via dispersion engineering \cite{Okawachi2014} and can reach up to octave-spanning coverage \cite{Pfeiffer2017}. In addition to their spectral properties, recent advances in fabrication technology have significantly reduced the power requirements for DKS generation, thus allowing for direct integration with semiconductor pump lasers \cite{Liu:18a, Stern:2018ca}. Altogether, through their exceptional optical properties and wafer-scale fabrication, DKS microcombs are promising candidates as sources for imaging and in particular OCT.
Here, we demonstrate for the first time the use of a soliton microcomb for SD-OCT. 

\begin{figure*}[!t]
\includegraphics[width = 0.7\textwidth]{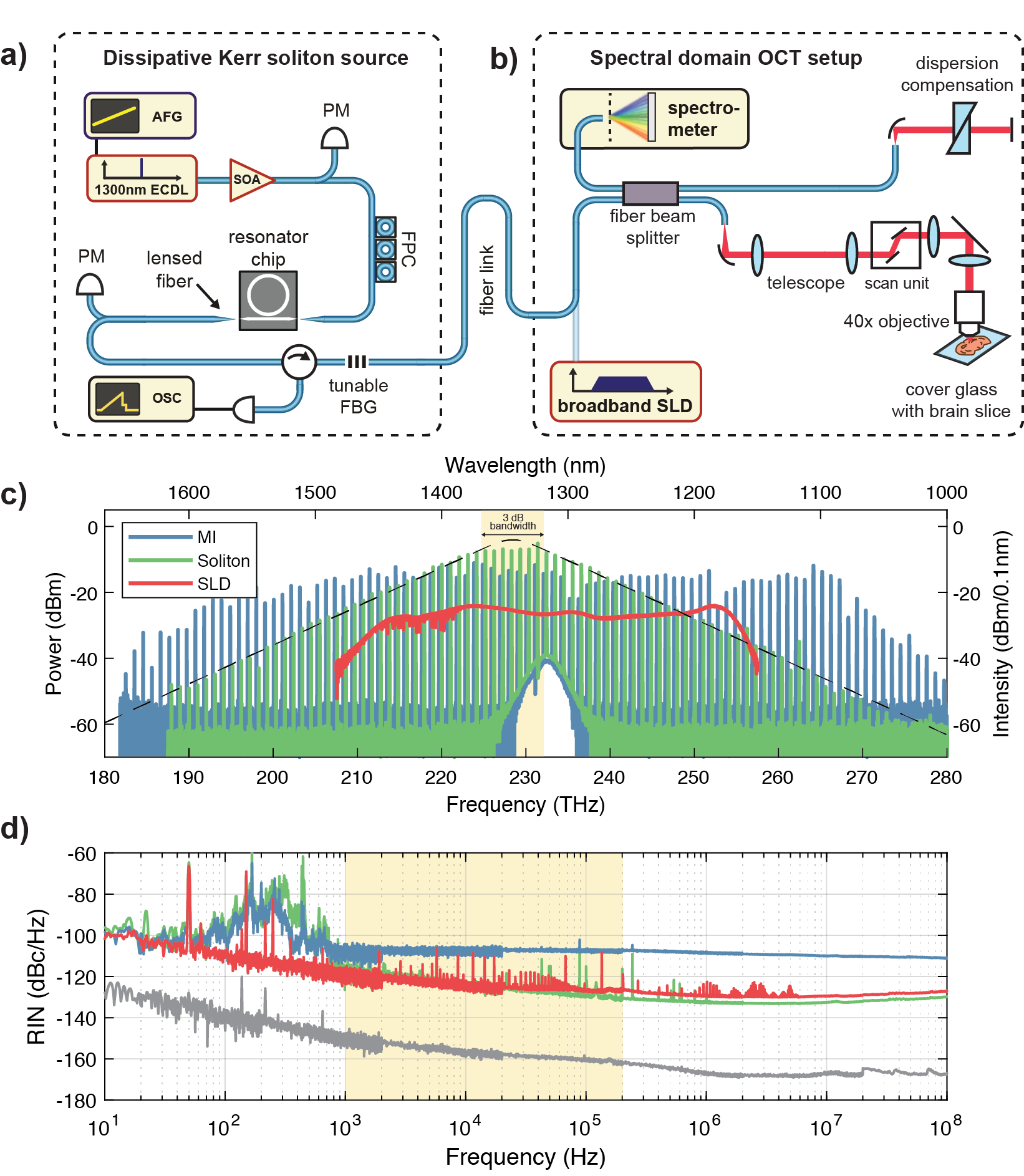}
\caption{\textbf{Experimental demonstration of DKS enabled SD-OCT.} a) Setup for DKS frequency comb generation based on a $1300\,\mathrm{nm}$ external cavity diode laser (ECDL) amplified by a semiconductor optical amplifier (SOA). The laser wavelength is tuned by a voltage ramp provided by the arbitrary function generator (AFG) and monitored by power meters (PM). After coupling to the chip using lensed fibers, the transmitted light intensity is displayed on an oscilloscope (OSC) and provides information about the tuning process. A tunable fiber Bragg grating (FBG) is used to suppress the pump light before sending the generated light over a fiber link to the OCT setup located in a different laboratory on the campus. b) SD-OCT setup based on a fiberized interferometer with a dispersion compensated reference arm and a high-resolution spectrometer. The setup was designed for use with a broadband SLD and the DKS comb signal was inserted without further modification for imaging. c) A chaotic modulation instability comb (blue) and a dissipative Kerr soliton (DKS) state (green) exhibiting spectral bandwidths comparable to the commercial SLD (orange). The DKS spectrum follows the characteristic $\mathrm{sech^2}$ profile and has a low density of avoided modal crossings. (d) The associated relative intensity noise (RIN) of the Kerr combs and the SLD. Note that the two Kerr comb states were generated in different resonators, as detailed in the Methods section. Here, the yellow shaded region represents the frequencies of interest for OCT measurements.}
\label{fig_combSource}
\end{figure*} 

\noindent\textbf{Dissipative Kerr solitons as a source for SD-OCT.}
We designed novel microcombs sources for OCT imaging operating in the second optical window (NIR-II), at 1300~nm, for its relatively low water absorption and reduced tissue scattering properties. We fabricated three \ce{Si3N4} resonators (one shown in the inlet of Fig. \ref{fig_octPrinciple} e)) following the established photonic Damascene process \cite{Pfeiffer2016} with free spectral range (FSR) of $\sim 100,\,\ 200 \,\ \mathrm{GHz}$ and $1 \,\ \mathrm{THz}$, respectively (Fig. \ref{fig_octPrinciple} e). Through their large waveguide cross sections, the resonators achieve anomalous group velocity dispersion (GVD) in the NIR-II imaging window (see Methods and S.I. for details). A microcomb, as shown in Fig. \ref{fig_octPrinciple} a), is generated by the nonlinear frequency conversion processes inside a micro-resonator \cite{Kippenberg2011}. The mutual interplay between (non-)degenerate four-wave mixing processes and self- and cross-phase modulations provides an optical gain to the resonator modes adjacent to the pumped mode. The Kerr comb generation is achieved by sweeping the pump laser frequency from the effective blue-detuned to a defined point at the effective red-detuned side of the selected cavity resonance. For DKS comb generation, the laser sweeping typically stops at a multi-soliton state and proceeds to a single soliton state through a backward frequency tuning technique \cite{Guo2016}.


As illustrated in Fig. \ref{fig_combSource}, the nonlinear frequency conversion bandwidth of the 1~THz microcombs can readily reach and exceed the bandwidth of SLDs. This is demonstrated for two distinctly different operational Kerr frequency comb states: the DKS and the chaotic modulation instability (MI) states (shown in Fig. \ref{fig_combSource} c). The DKS state, shown in green, exhibits a characteristic $\mathrm{sech}^2$ spectral envelope and reaches a spectral coverage similar to the reference SLD source. The cross section of the 1~THz DKS waveguide, $780 \times 1450\,\mathrm{nm^2}$, provides an anomalous GVD ($ \mathrm{D_2/2\pi} \sim 40\,  \mathrm{MHz}$) for soliton pulse formation. The 3~dB bandwidth of the DKS spectrum, highlighted in Fig. \ref{fig_combSource} c), is $\sim 8.3\,\mathrm{THz}$, corresponding to a 38~fs transform limited pulse.

The chaotic Kerr comb state, shown in blue in Fig. \ref{fig_combSource} c), provides a spectral coverage well exceeding the SLD's, due to the lower GVD ($\mathrm{D_2/2\pi} \sim 20\,\mathrm{MHz}$) originating from its smaller micro-resonator cross section ($730 \times 1425\,\mathrm{nm^2}$). The resulting spectral envelope is overall flat but, in contrast to the DKS state, exhibits local power variations caused by avoided mode crossings. \\ 


\begin{figure*}[!t]
\includegraphics[width = 0.95\textwidth]{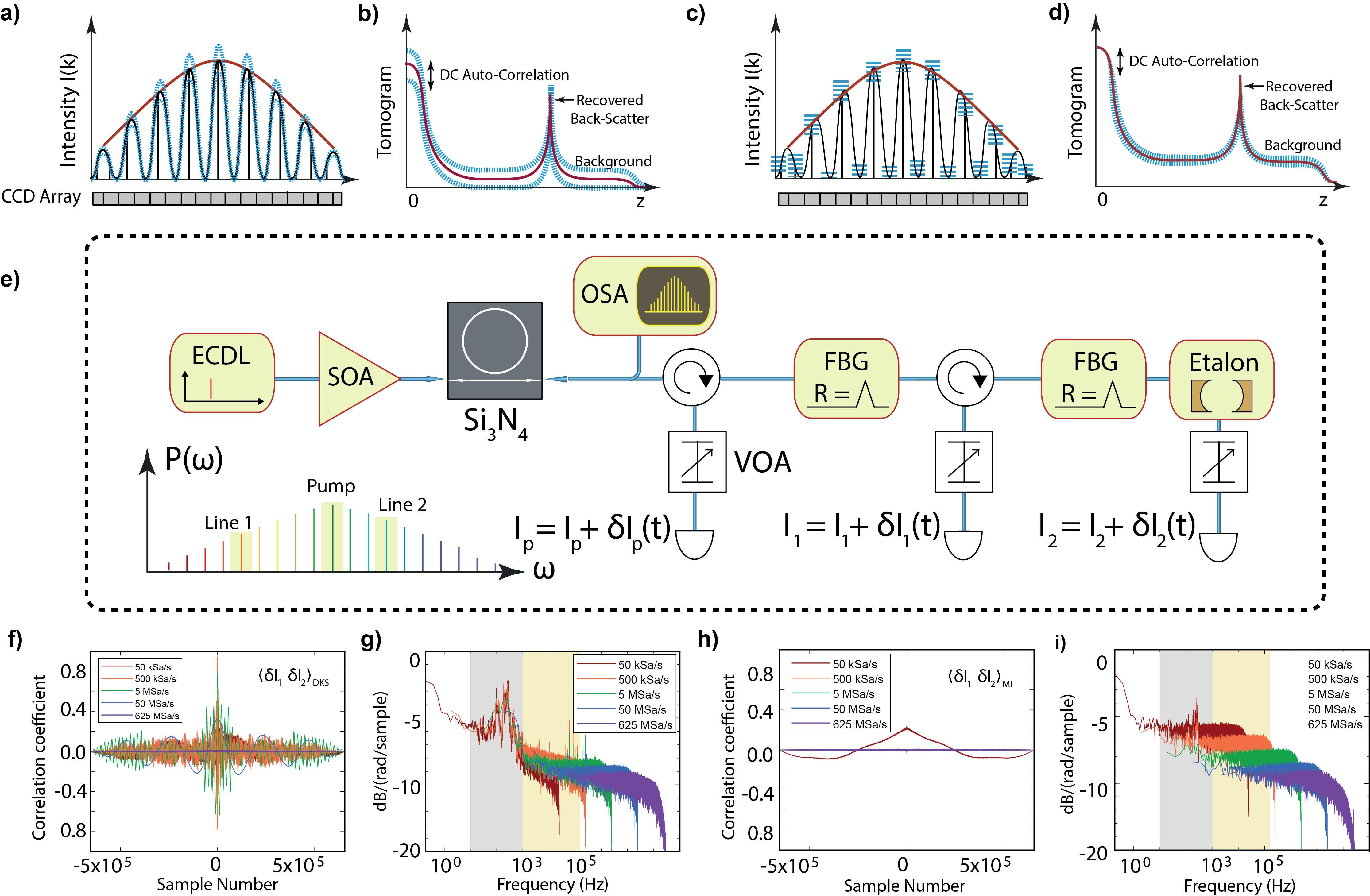}
\caption{\textbf{ Frequency Dependent Noise Correlations} a) A frequency comb interferogram (black), which is dispersively projected onto a CCD array. Correlated intensity noise (blue) modulates the full comb envelope (red). b) Tomogram corresponding to the spectrum in a). The DC peak is sensitive to the noise, but no change in background SNR and dynamic range occurs. c) Uncorrelated intensity noise between various comb lines, which manifests in the interferogram seen in d) as an increase in background signal. e) Setup for intensity noise correlation measurement. DKS and MI states are generated by laser piezo tuning. Various lines are then filtered from the resulting spectrum, sent through variable optical attenuators (VOA), and sampled with a high-resolution oscilloscope. d) and f) The cross correlation of lines "1" and "2" for the DKS and MI states, where lines "1" and "2" correspond to 1272 nm and 1320 nm, respectively. The x-axis denotes the relative lag in units of the sampling rate, as derived from the samples per second (Sa/s), with all color indications shared between sub-figures. f) and h) The cross power spectral densities corresponding to g) and i). Here, the grey shaded region represents technical noise, likely originating from acoustic modes of the input- and output coupling fibers, while the yellow shaded region indicates the range of typical SD-OCT A-scan rates.}
\label{fig_xcorr}
\end{figure*} 

\noindent\textbf{Noise characteristics of soliton microcombs.}
To assess the noise characteristics of these novel sources and their applicability to OCT imaging, we first measure their relative intensity noise $ \mathrm{RIN} = \frac{S_{P}(f)}{\langle P^2\rangle}$, with $S_{P}(f)$ denoting the single sided power spectral density of the intensity fluctuations (shown in Fig.~\ref{fig_combSource}~d), and demonstrate that while the MI state provides a broader spectral coverage, its chaotic nature results in an increase in RIN of nearly 20~dB, extending to very high offset frequencies in the GHz domain (Fig.~\ref{fig_combSource}~d) \cite{herr_universal_2012}. These measurements were performed for different FSRs (i.e. $100$, $200$ $\mathrm{GHz}$ and $1$ $\mathrm{THz}$), and resulted in similar RIN profiles between resonators (data not shown here). Accordingly, although chaotic comb states in a $ \mathrm{Si_3N_4}$ microresonator have been demonstrated in OCT imaging \cite{Ji:19}, their higher noise should ultimately limit OCT performance as compared to SLDs, especially at elevated imaging speeds. Meanwhile, we also show that the DKS soliton state has comparable intensity noise with the SLD, at frequencies higher than 10~kHz. In the low frequency regime, mechanical modes of the input and output lensed fiber-coupling result in a broad noise peak spanning from 100 to 1000~Hz for both the MI and DKS states, which can be mitigated through optimized packaging or feedback loops.

Even more so, the ultimate performance limit of coherent sources at high offset frequencies, such as the DKS comb, is given by the photon shot noise (RIN = $\frac{2\hbar\omega}{P} = \mathrm{-145~dBc/Hz}$ with 20~$\mu$W power on the detector) and improves with optical power. In contrast, in the case of broadband, incoherent light sources, the RIN is limited by spontaneous emission beat noise \cite{yurek1986quantum,dericksen13} ($\mathrm{RIN} = 1/B_0 = \mathrm{-136~dBc/Hz}$ for a 45~THz rectangular bandwidth SLD source), which ultimately limits the dynamic range gain with high source powers in the reference arm \cite{sorin1992simple}. 

Next, we explore one unique feature of soliton microcombs; the high-degree of coherence between individual comb lines. This is especially important in the context of OCT, as line-by-line intensity noise of the frequency comb's retrieved spectra (Fig. \ref{fig_xcorr} a) and c)) corresponds to pixel-by-pixel noise in the retrieved image (Fig. \ref{fig_xcorr} b) and d)). Indeed, as an image in SD-OCT is produced via a Fourier transform of the interferogram, only uncorrelated intensity noise between various pixels degrades the final image \cite{yun2004pulsed}. Noise in the amplitude of the spectrum's enveloppe will be act only on the DC component of the tomogram (Fig. \ref{fig_xcorr} a) and b), whereas uncorrelated intensity fluctuations between the different optical frequencies will lead to a higher noise level at all depths of the tomogram (Fig. \ref{fig_xcorr} c) and d)). To investigate these intra-tone noise properties, we performed the cross-correlation of intensity fluctuations on pairs of comb lines, using the experimental setup described in Fig. \ref{fig_xcorr} e) \cite{siegman1971introduction}. From both DKS or MI combs, individual comb lines are filtered and time traces are recorded and cross-correlated (Fig.~\ref{fig_xcorr}~f) and g) for various sampling speeds. The corresponding cross power spectral densities (PSD) are depicted in Figures \ref{fig_xcorr} g) and i). In practice, we chose two lines, at 1272~$\mathrm{nm}$ and at 1320~$\mathrm{nm}$ (lines~1 and~2 respectively in Fig.~\ref{fig_xcorr}). In the DKS state, we observe a peak correlation coefficient between the two chosen lines of approximately~0.95, corresponding to a sampling rate of $\mathrm{500 \,\ kSa/s}$. The maximum correlation coefficient near zero lag stays well above~0.8 for sampling frequencies up to $\mathrm{5\,\ MSa/s}$, indicating that intensity noise between DKS comb lines is highly correlated even at elevated frequencies. In contrast, for the fully developed MI state, the maximum correlation coefficient between lines~1 and~2 is approximately~0.24, and occurs for the lowest sampling speed (DC). For all higher sampling frequencies, however, the correlation coefficient decreases to approximately 0.01, indicating highly uncorrelated intensity noise between comb lines. We expect a similar behavior for the the classical noise of nearly all incoherent sources, including for SLD sources.

As mentioned earlier, given that the ultimate limit of the noise properties of frequency domain OCT is set by the degree of correlation of intensity noise between various spectral channels \cite{yun2004pulsed}, and therefore different optical frequencies, the DKS state can offer significant advantages, in terms of noise, as compared to the MI state. In view of these differences in noise performances, as well as the DKS's superior nonlinear efficiency and bandwidth, we chose to use a DKS source for the OCT experiments presented here. \\

\noindent\textbf{Spectral characteristics of microcombs and their implications for OCT imaging and circular ranging.} In frequency domain OCT, depth-resolved information about the sample is conveyed through the amplitude and frequency of an inteferogram. A reflectivity profile is obtained through a Fourier transform of the recorded spectrum on the spectrometer. From sampling theory, the maximum imaging depth obtainable $z_\mathrm{max}$ is therefore dictated by the spectrometer's spectral resolution $\delta k_{\mathrm{CCD}}$ as \cite{Izatt2008}:
\begin{equation}
    \pm z_\mathrm{max} = \pm \frac{1}{4 \delta k_{\mathrm{CCD}}}
        \label{eq_zmax}
\end{equation}
As such, OCT systems designed for high axial resolution and deep penetration imaging require a detection capable of registering a broadband spectra at a fine spectral resolution. In practice, combining these two features is cumbersome in SD-OCT due to the limited length of current array detectors (typically between 1024 and 2048, and exceptionally 8196~pixels \cite{Lichtenegger2018}), ultimately limiting either the effective resolution or the available imaging range.

When comb-like sources, such as Kerr combs, are employed instead of a continuous spectrum, the discrete set of frequencies will generate a periodicity in the tomogram if the frequency/time difference between the combs is sampled by the detector \cite{Siddiqui2018}. The frequency of this periodicity, called the ambiguity range, is determined by the source's repetition rate $f_\mathrm{rep}$ (which also corresponds to the temporal separation between the individual pulses). For Kerr combs, the repetition rate is given by the micro-resonator FSR~($\mathrm{D_1/2\pi}$):

\begin{equation}
    z_\mathrm{ambiguity} = \frac{c}{2n_\mathrm{tissue}}\frac{1}{f_\mathrm{rep}} \approx \frac{c}{2n_\mathrm{tissue}}\frac{2\pi}{\mathrm{D_1}}
\end{equation}

with the speed of light \textit{c} and the tissue refractive index $n_\mathrm{tissue}$. For the imaging experiments carried out below, we used micro-resonators with a $1\,\mathrm{THz}$ FSR, leading to an ambiguity range of $\sim71\,\mathrm{\mu m}$ compared to a maximum imaging range of $\sim 2\,\mathrm{mm}$ offered by the spectrometer. In contrast, the lower FSR DKS sources shown in Fig. \ref{fig_octPrinciple} e) offer repetition rates down to 100~GHz, corresponding to an increased ambiguity range of $\sim710\,\mathrm{\mu m}$.\

In addition to their discreteness in frequency, DKS sources also possess interesting temporal coherence properties. Although the overall coherence length of the source is dictated by its bandwidth, the coherence length of each comb tone of the DKS source equals that of the driving pump laser and thus amounts to several kilometers for a pumping linewidth around $100\,\mathrm{kHz}$. 
As mentioned earlier and highlighted in Equation \ref{eq_zmax}, the attainable imaging range in FD-OCT is typically dictated either by the spectral resolution of the spectrometer or by the width of the swept spectral line (for spectral-domain and swept-source respectively). When combining DKS sources with an SD-OCT system, a mismatch can therefore occur between the imaging range (given by the spectrometer, here $\sim 2$~mm) and the coherence length of each comb tone (here~$> 2$~km). As such, the coherence lengths reached here largely exceed the imaging ranges of typical OCT systems, entailing novel advantages and disadvantages for imaging, which will be discussed below.\\

\begin{figure*}[!t]
\includegraphics[width = \textwidth]{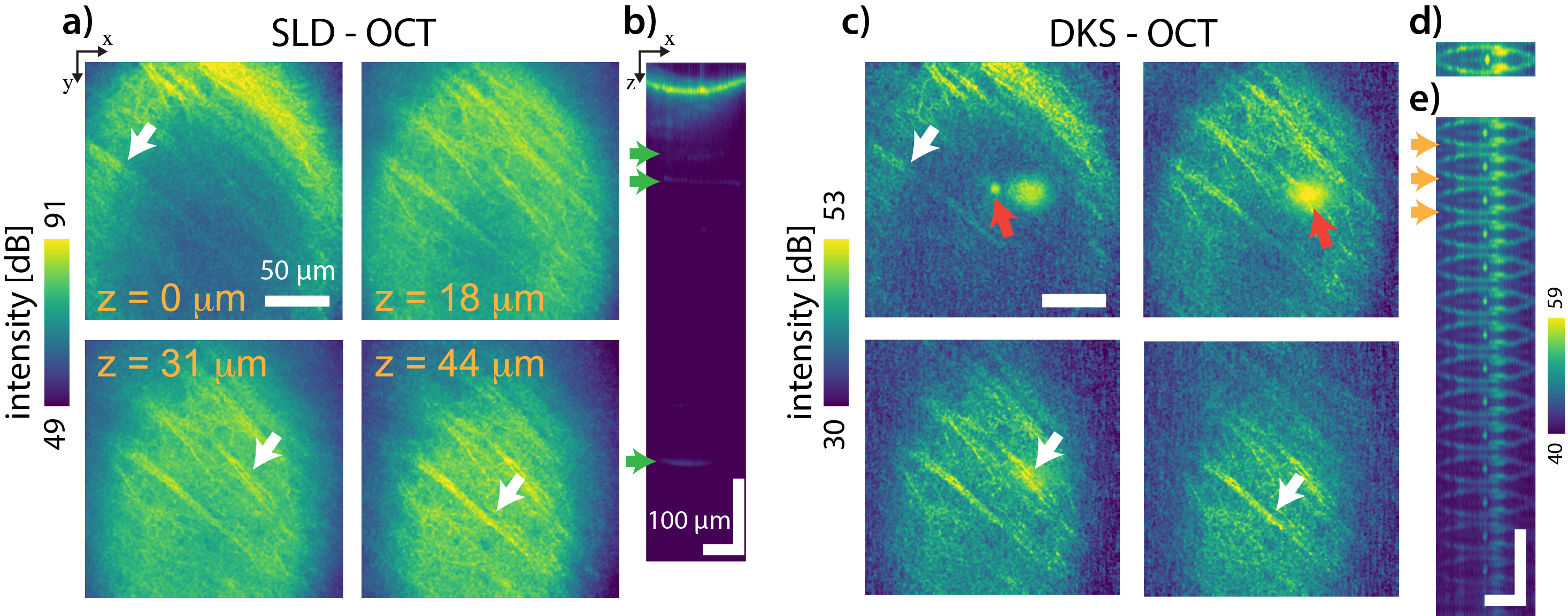}
\caption{\textbf{Qualitative performance comparison of SLD and DKS OCT for \textit{ex vivo} cerebral tissue imaging.} En-face images at different depths of a slice of brain tissue were obtained with both SLD and DKS sources a) and c) respectively, revealing the presence of highly scattering neuronal fibers (pointed by blue arrows). The en-face views obtained with the DKS source also contain additional features, pointed by red arrows, such as bright vertical stripes, circular ring patterns, and higher intensity regions. The cross-sections for both SLD and DKS imaging b) and d) respectively, highlight the imaging's field curvature, highly reflective structures below the sample (pointed by white arrows) and the presence of an ambiguity range when imaging with the discrete DKS source (pointed by red arrows).}
\label{fig_imagingResults}
\end{figure*} 

\noindent\textbf{OCT imaging with a DKS microcomb.}
The difference in performance between the SLD and the DKS as sources for OCT imaging was qualitatively assessed by imaging a $\sim 50 \,\ \mu$m thick slice of a mouse brain tissue. The OCT was equipped with a 40$\times$ 0.8 NA objective (Olympus) to obtain a lateral resolution of $\sim 1.5 \,\ \mu$m and a depth-of-field shorter than the source's ambiguity range. In a first step, we imaged the slice using the SLD source, providing an axial resolution of $\sim$ 6 $\mu$m in air. Figure \ref{fig_imagingResults} a) presents en-face views over a $200 \times 200 \,\ \mu$m$^2$ area at specific depths, whereas panel b) shows the cross-section images of the SLD based OCT tomogram. These views present similar features as other OCT observations of cerebral tissues \cite{Srinivasan2012a}, such as neural fibers (pointed by white arrows), which appear as directional, bright, and fine structures over dim neuropil. Within the neuropil, darker circular structures seemingly point to the presence of neuronal cell bodies, as already observed in high resolution OCT \cite{Srinivasan2012a, Assayag2013}. 

Secondly, without modifying any imaging parameters nor touching the sample, the SLD was disconnected from the system and replaced with the DKS source, providing an axial resolution of $\sim$ 10 $\mu$m in air. Figures \ref{fig_imagingResults} c), d) and e) show the OCT tomogram of the same sample with the DKS light source. The neural fibers can be clearly observed in the en-face views with higher contrast. However, the neuropil appears darker and fewer details can be discerned. Additional artifacts, as indicated by the red arrows, are present in some of the DKS views and are likely caused by the combination of two characteristics of the DKS source: its discrete set of frequencies and its narrow linewidth. Overall, the dynamic range obtained in the DKS images is reduced by $\sim$ 19~dB compared to the SLD. This discrepancy could originate from the significantly lower power provided by the DKS source (estimated to be up to a fourth of the SLD power) and from the presence of the spurious back-reflections, ultimately drowning the collection of weakly scattering features. For both sources, the A-scan rate was maintained at 46 kHz. The images presented in panels c) and d) were obtained by selecting only the comb tones from the interferograms, dismissing non-illuminated pixels. Conversely, for panel~f) the entire recorded interferogram was used. More details on the processing are available in the S.I. The axial resolution of the SLD and the DKS were extracted using a reflective mirror (as shown in S.I. Fig. \ref{fig_mirror}) and are $\sim$ 6 $\mu$m and $\sim$ 10 $\mu$m respectively.

As mentioned earlier, the frequency discretization of the source will lead to a periodic image folding along the axial dimension, as similarly observed by Siddiqui et al. \cite{Siddiqui2018}. The ambiguity range of the source can be observed in the cross-section (Fig. \ref{fig_imagingResults} e) and manifests itself as an axial periodicity of the structures (orange arrows in Fig. \ref{fig_imagingResults} e). As the comb width is significantly narrower than the spectrometer's spectral resolution, the coherence length of the DKS comb tones exceeds both the ambiguity range and the spectrometer's imaging range (Fig. \ref{fig_octPrinciple} b). The aforementioned image folding and extended coherence length thus allows reflections within the optical path to interfere with the reference arm, and will ultimately be superimposed with the features under investigation. As a result, some of the artifacts in the DKS images might stem from the folding of structures beyond the DKS's ambiguity range, such as reflections from optical components and the coverslide (illustrated in Fig. \ref{fig_octPrinciple} a) or from the back-scattering of cerebral structures. Some of the artefacts pointed by red arrows in Fig. \ref{fig_imagingResults} c) can be observed at deeper locations in the SLD's tomogram, highlighted by green arrows in Fig. \ref{fig_imagingResults} b). Typically, these strong reflections will occupy a significant portion of the spectrometer's dynamic range and could ultimately drown the fine details of the image, as previously observed in OCT \cite{Villiger:10,Blatter2011}.\\




\noindent\textbf{Future direction of the DKS based SD-OCT.}
In this manuscript, we have demonstrated for the first time the use of a DKS source for SD-OCT. We show that such soliton sources (DKS) are an interesting candidate for SD-OCT imaging through their low-noise, discrete set of frequencies and large bandwidths. Our work highlights the high noise performance of the source: specifically, the DKS (a coherent broadband source), equals and even outperforms an SLD (fully incoherent source) in its relative intensity noise (RIN). Equally important, DKS feature a unique property, in our knowledge previously unseen in OCT sources: the noise between the comb tones comprising the soliton frequency combs shows an unprecedentedly high degree of correlation. This feature is particularly important in OCT, as images are obtained through a Fourier transform of the spectrum. As such, noise common to all comb tones does not degrade the dynamic range, whereas relative uncorrelated fluctuations from pixel to pixel contribute to a significant dynamic range reduction \cite{yun2004pulsed}. With the noise of the source characterized, we imaged \emph{ex vivo} mice fixed brain slices, and found that the novel source allows for visualization of similar features to an SLD source, although with an overall reduced dynamic and imaging range. Overcoming these pitfalls can be achieved by optimizing both the OCT instrument and the source. First, the artefacts present in Fig. \ref{fig_imagingResults} c)-e) could be suppressed either by using solely reflective optical elements \cite{Amirsolaimani:17} or through a dark-field implementation \cite{Villiger:10,Blatter2011}. As these spurious reflections can occupy a significant portion of the dynamic range of the camera, eliminating these features could help further enhance the system's imaging capabilities. Second, the $\sim71\,\mathrm{\mu m}$ ambiguity range available with 1~THz DKS is too short for most imaging applications. It is however sufficient for imaging of thin flat tissues and for certain optical biopsy applications \cite{belykh2018utilization,sanai2011intraoperative,schlosser2011confocal,charalampaki2015confocal,zehri2014neurosurgical,fugazza2016confocal,wellikoff2015comparison,bui2015pilot,fuks2018intraoperative,krafft2018perspectives,belykh2018probe,lombardini2018high}, wherein there is need for a real-time assessment of brain and tumor tissue on a cellular level, as patient survival has been shown to be correlated to the extent of tumor resection \cite{stummer2006fluorescence}. Moreover, as shown in Fig. \ref{fig_octPrinciple} e), DKS sources with shorter FSRs down to 100 and 200~GHz are already available with similar noise profiles as the one used here for imaging. These sources would enable reaching ambiguity ranges up to $\sim 0.7$~mm, which are compatible with most \emph{in vivo} imaging applications\cite{Siddiqui2018}. Third, fully exploiting the circular ranging capabilities of the source requires reading the interferograms in a complex-valued form \cite{Siddiqui2012}, which can be attained by adding acousto-optic frequency shifters to the system \cite{Lippok2018, Bachmann:06}. Lastly, the central wavelength of 1300~nm used here is not suitable for all \emph{in vivo} applications, especially human ophtalmology. Nevertheless, the source's design can be modified, enabling shifting of the central wavelength to shorter spectral ranges, such as 1~$\mu$m, as demonstrated previously \cite{karpov_photonic_2018,lee2017towards}.

Overall, in addition to the unprecedented noise performance of the DKS source and the increased imaging efficiency available through optical-domain subsampling, frequency combs could potentially alleviate certain shortcomings of SD-OCT detection schemes by facilitating $\lambda$-to-k mapping and reducing depth dependant sensitivity roll-off \cite{Bajraszewski2008a, Tsai2009}. DKS sources could also lead to higher axial resolutions at 1300~nm: as highlighted in Fig. \ref{fig_combSource} c), the power spectral density of the DKS source exceeds the SLD's from $\sim$ 1250~nm to 1500~nm. As such, using spectral shaping, the DKS could provide a bandwidth comparable or larger than current broadband SLDs used for 1300~nm imaging.

The high performance of the DKS source could lead to a significant miniaturization of the OCT system. The optical-domain sub-sampling capabilities of the source, highlighted in Fig. \ref{fig_imagingResults} c), already indicate a potential shortening of the reference arm of $\sim 2$ mm. Furthermore, although not demonstrated here, the long coherence length of the DKS combs could enable further shortening of the length of the reference arm, reducing the instrument's footprint. In traditional SD-OCT systems, the path delay difference between the reference and sample arms needs to be smaller than the maximum imaging range of the spectrometer to record an interference. In the case of a frequency comb, this condition is alleviated through optical sub-sampling, so long as the path delay difference is within the coherence length of each line of the source. As the DKS source used in this study has a theoretical coherence length for each comb line beyond a kilometer, the reference arm length could be significantly shortened, ultimately paving the way to future miniaturized and potentially more efficient high-resolution OCT imaging systems. Lastly, the optical-domain subsampling properties of our source would be highly valuable in human \textit{in vivo} imaging, wherein the sample geometry is often non-planar and features could exceed the imaging range, such as in ophtalmology and intra-operative OCT. Altogether, the aforementioned noise and spectral properties of DKS microcombs hint to their significant unexplored potential for future exploitation in SD-OCT.

\section*{Data availability statement}
The data and code used to produce the results of this manuscript will be available on \texttt{Zenodo} upon publication.
\medskip

\section*{Authors contributions}
P.J.M. and M.H.P.P. performed the experiments. P.J.M., J.-J.H., M.H.H.P. analyzed the data. J.L. designed and fabricated the \ce{Si3N4} chip devices. P.J.M. designed and constructed the OCT setup. T.J.K., T.L., and C.H. supervised the project. J.C.S and J.R performed and analyzed all noise-related measurements. All of the authors contributed to the manuscript. 
\subsection*{Acknowledgements}
This work was supported by an industrial grant with Carl Zeiss AG "Optical coherence tomography with chip-scale Kerr soliton frequency combs". \ce{Si3N4} micro-resonator samples were fabricated in the EPFL Center of MicroNanotechnology (CMi). This publication was supported by the Swiss National Science Foundation under grant agreement No. 165933. P.J.M. acknowledges that part of his contribution was undertaken thanks to funding from the Canada First Research Excellence Fund through the TransMedTech Institute and from the Swiss National Science Foundation.

\bibliography{kerrOCTbibliography}

\begin{thebibliography}{65}%
\makeatletter
\providecommand \@ifxundefined [1]{%
 \@ifx{#1\undefined}
}%
\providecommand \@ifnum [1]{%
 \ifnum #1\expandafter \@firstoftwo
 \else \expandafter \@secondoftwo
 \fi
}%
\providecommand \@ifx [1]{%
 \ifx #1\expandafter \@firstoftwo
 \else \expandafter \@secondoftwo
 \fi
}%
\providecommand \natexlab [1]{#1}%
\providecommand \enquote  [1]{``#1''}%
\providecommand \bibnamefont  [1]{#1}%
\providecommand \bibfnamefont [1]{#1}%
\providecommand \citenamefont [1]{#1}%
\providecommand \href@noop [0]{\@secondoftwo}%
\providecommand \href [0]{\begingroup \@sanitize@url \@href}%
\providecommand \@href[1]{\@@startlink{#1}\@@href}%
\providecommand \@@href[1]{\endgroup#1\@@endlink}%
\providecommand \@sanitize@url [0]{\catcode `\\12\catcode `\$12\catcode
  `\&12\catcode `\#12\catcode `\^12\catcode `\_12\catcode `\%12\relax}%
\providecommand \@@startlink[1]{}%
\providecommand \@@endlink[0]{}%
\providecommand \url  [0]{\begingroup\@sanitize@url \@url }%
\providecommand \@url [1]{\endgroup\@href {#1}{\urlprefix }}%
\providecommand \urlprefix  [0]{URL }%
\providecommand \Eprint [0]{\href }%
\providecommand \doibase [0]{http://dx.doi.org/}%
\providecommand \selectlanguage [0]{\@gobble}%
\providecommand \bibinfo  [0]{\@secondoftwo}%
\providecommand \bibfield  [0]{\@secondoftwo}%
\providecommand \translation [1]{[#1]}%
\providecommand \BibitemOpen [0]{}%
\providecommand \bibitemStop [0]{}%
\providecommand \bibitemNoStop [0]{.\EOS\space}%
\providecommand \EOS [0]{\spacefactor3000\relax}%
\providecommand \BibitemShut  [1]{\csname bibitem#1\endcsname}%
\let\auto@bib@innerbib\@empty
\bibitem [{\citenamefont {Drexler}\ and\ \citenamefont
  {Fujimoto}(2015)}]{Drexler2015}%
  \BibitemOpen
  \bibfield  {author} {\bibinfo {author} {\bibfnamefont {W.}~\bibnamefont
  {Drexler}}\ and\ \bibinfo {author} {\bibfnamefont {J.~G.}\ \bibnamefont
  {Fujimoto}},\ }\href@noop {} {\emph {\bibinfo {title} {{Optical Coherence
  Tomography: Technology and Applications}}}},\ \bibinfo {edition} {2nd}\ ed.\
  (\bibinfo  {publisher} {Springer},\ \bibinfo {year} {2015})\BibitemShut
  {NoStop}%
\bibitem [{\citenamefont {Kippenberg}\ \emph {et~al.}(2018)\citenamefont
  {Kippenberg}, \citenamefont {Gaeta}, \citenamefont {Lipson},\ and\
  \citenamefont {Gorodetsky}}]{kippenberg2018dissipative}%
  \BibitemOpen
  \bibfield  {author} {\bibinfo {author} {\bibfnamefont {T.~J.}\ \bibnamefont
  {Kippenberg}}, \bibinfo {author} {\bibfnamefont {A.~L.}\ \bibnamefont
  {Gaeta}}, \bibinfo {author} {\bibfnamefont {M.}~\bibnamefont {Lipson}}, \
  and\ \bibinfo {author} {\bibfnamefont {M.~L.}\ \bibnamefont {Gorodetsky}},\
  }\href@noop {} {\bibfield  {journal} {\bibinfo  {journal} {Science}\ }\textbf
  {\bibinfo {volume} {361}},\ \bibinfo {pages} {eaan8083} (\bibinfo {year}
  {2018})}\BibitemShut {NoStop}%
\bibitem [{\citenamefont {Kues}\ \emph {et~al.}(2019)\citenamefont {Kues},
  \citenamefont {Reimer}, \citenamefont {Lukens}, \citenamefont {Munro},
  \citenamefont {Weiner}, \citenamefont {Moss},\ and\ \citenamefont
  {Morandotti}}]{kues2019quantum}%
  \BibitemOpen
  \bibfield  {author} {\bibinfo {author} {\bibfnamefont {M.}~\bibnamefont
  {Kues}}, \bibinfo {author} {\bibfnamefont {C.}~\bibnamefont {Reimer}},
  \bibinfo {author} {\bibfnamefont {J.~M.}\ \bibnamefont {Lukens}}, \bibinfo
  {author} {\bibfnamefont {W.~J.}\ \bibnamefont {Munro}}, \bibinfo {author}
  {\bibfnamefont {A.~M.}\ \bibnamefont {Weiner}}, \bibinfo {author}
  {\bibfnamefont {D.~J.}\ \bibnamefont {Moss}}, \ and\ \bibinfo {author}
  {\bibfnamefont {R.}~\bibnamefont {Morandotti}},\ }\href@noop {} {\bibfield
  {journal} {\bibinfo  {journal} {Nature Photonics}\ }\textbf {\bibinfo
  {volume} {13}},\ \bibinfo {pages} {170} (\bibinfo {year} {2019})}\BibitemShut
  {NoStop}%
\bibitem [{\citenamefont {Ji}\ \emph {et~al.}(2019)\citenamefont {Ji},
  \citenamefont {Yao}, \citenamefont {Klenner}, \citenamefont {Gan},
  \citenamefont {Gaeta}, \citenamefont {Hendon},\ and\ \citenamefont
  {Lipson}}]{Ji:19}%
  \BibitemOpen
  \bibfield  {author} {\bibinfo {author} {\bibfnamefont {X.}~\bibnamefont
  {Ji}}, \bibinfo {author} {\bibfnamefont {X.}~\bibnamefont {Yao}}, \bibinfo
  {author} {\bibfnamefont {A.}~\bibnamefont {Klenner}}, \bibinfo {author}
  {\bibfnamefont {Y.}~\bibnamefont {Gan}}, \bibinfo {author} {\bibfnamefont
  {A.~L.}\ \bibnamefont {Gaeta}}, \bibinfo {author} {\bibfnamefont {C.~P.}\
  \bibnamefont {Hendon}}, \ and\ \bibinfo {author} {\bibfnamefont
  {M.}~\bibnamefont {Lipson}},\ }\href {\doibase 10.1364/OE.27.019896}
  {\bibfield  {journal} {\bibinfo  {journal} {Opt. Express}\ }\textbf {\bibinfo
  {volume} {27}},\ \bibinfo {pages} {19896} (\bibinfo {year}
  {2019})}\BibitemShut {NoStop}%
\bibitem [{\citenamefont {Siddiqui}\ and\ \citenamefont
  {Vakoc}(2012)}]{Siddiqui2012}%
  \BibitemOpen
  \bibfield  {author} {\bibinfo {author} {\bibfnamefont {M.}~\bibnamefont
  {Siddiqui}}\ and\ \bibinfo {author} {\bibfnamefont {B.~J.}\ \bibnamefont
  {Vakoc}},\ }\href {\doibase 10.1364/OE.20.017938} {\bibfield  {journal}
  {\bibinfo  {journal} {Optics Express}\ }\textbf {\bibinfo {volume} {20}},\
  \bibinfo {pages} {17938} (\bibinfo {year} {2012})}\BibitemShut {NoStop}%
\bibitem [{\citenamefont {Siddiqui}\ \emph {et~al.}(2018)\citenamefont
  {Siddiqui}, \citenamefont {Nam}, \citenamefont {Tozburun}, \citenamefont
  {Lippok}, \citenamefont {Blatter},\ and\ \citenamefont
  {Vakoc}}]{Siddiqui2018}%
  \BibitemOpen
  \bibfield  {author} {\bibinfo {author} {\bibfnamefont {M.}~\bibnamefont
  {Siddiqui}}, \bibinfo {author} {\bibfnamefont {A.~S.}\ \bibnamefont {Nam}},
  \bibinfo {author} {\bibfnamefont {S.}~\bibnamefont {Tozburun}}, \bibinfo
  {author} {\bibfnamefont {N.}~\bibnamefont {Lippok}}, \bibinfo {author}
  {\bibfnamefont {C.}~\bibnamefont {Blatter}}, \ and\ \bibinfo {author}
  {\bibfnamefont {B.~J.}\ \bibnamefont {Vakoc}},\ }\href {\doibase
  10.1038/s41566-017-0088-x} {\bibfield  {journal} {\bibinfo  {journal} {Nature
  Photonics}\ }\textbf {\bibinfo {volume} {12}},\ \bibinfo {pages} {111}
  (\bibinfo {year} {2018})}\BibitemShut {NoStop}%
\bibitem [{\citenamefont {Huang}\ \emph {et~al.}(1991)\citenamefont {Huang},
  \citenamefont {Swanson}, \citenamefont {Lin}, \citenamefont {Schuman},
  \citenamefont {Stinson}, \citenamefont {Chang}, \citenamefont {Hee},
  \citenamefont {Flotte}, \citenamefont {Gregory}, \citenamefont {Puliafito},\
  and\ \citenamefont {et}}]{Huang1178}%
  \BibitemOpen
  \bibfield  {author} {\bibinfo {author} {\bibfnamefont {D.}~\bibnamefont
  {Huang}}, \bibinfo {author} {\bibfnamefont {E.}~\bibnamefont {Swanson}},
  \bibinfo {author} {\bibfnamefont {C.}~\bibnamefont {Lin}}, \bibinfo {author}
  {\bibfnamefont {J.}~\bibnamefont {Schuman}}, \bibinfo {author} {\bibfnamefont
  {W.}~\bibnamefont {Stinson}}, \bibinfo {author} {\bibfnamefont
  {W.}~\bibnamefont {Chang}}, \bibinfo {author} {\bibfnamefont
  {M.}~\bibnamefont {Hee}}, \bibinfo {author} {\bibfnamefont {T.}~\bibnamefont
  {Flotte}}, \bibinfo {author} {\bibfnamefont {K.}~\bibnamefont {Gregory}},
  \bibinfo {author} {\bibfnamefont {C.}~\bibnamefont {Puliafito}}, \ and\
  \bibinfo {author} {\bibfnamefont {a.}~\bibnamefont {et}},\ }\href {\doibase
  10.1126/science.1957169} {\bibfield  {journal} {\bibinfo  {journal}
  {Science}\ }\textbf {\bibinfo {volume} {254}},\ \bibinfo {pages} {1178}
  (\bibinfo {year} {1991})}\BibitemShut {NoStop}%
\bibitem [{\citenamefont {Fercher}\ \emph {et~al.}(2003)\citenamefont
  {Fercher}, \citenamefont {Drexler}, \citenamefont {Hitzenberger},\ and\
  \citenamefont {Lasser}}]{Fercher2003}%
  \BibitemOpen
  \bibfield  {author} {\bibinfo {author} {\bibfnamefont {A.~F.}\ \bibnamefont
  {Fercher}}, \bibinfo {author} {\bibfnamefont {W.}~\bibnamefont {Drexler}},
  \bibinfo {author} {\bibfnamefont {C.~K.}\ \bibnamefont {Hitzenberger}}, \
  and\ \bibinfo {author} {\bibfnamefont {T.}~\bibnamefont {Lasser}},\
  }\href@noop {} {\bibfield  {journal} {\bibinfo  {journal} {Rep. Prog. Phys.}\
  }\textbf {\bibinfo {volume} {66}},\ \bibinfo {pages} {239} (\bibinfo {year}
  {2003})}\BibitemShut {NoStop}%
\bibitem [{\citenamefont {Kassani}\ \emph {et~al.}(2017)\citenamefont
  {Kassani}, \citenamefont {Villiger}, \citenamefont {Uribe-Patarroyo},
  \citenamefont {Jun}, \citenamefont {Khazaeinezhad}, \citenamefont {Lippok},\
  and\ \citenamefont {Bouma}}]{Kassani2017}%
  \BibitemOpen
  \bibfield  {author} {\bibinfo {author} {\bibfnamefont {S.~H.}\ \bibnamefont
  {Kassani}}, \bibinfo {author} {\bibfnamefont {M.}~\bibnamefont {Villiger}},
  \bibinfo {author} {\bibfnamefont {N.}~\bibnamefont {Uribe-Patarroyo}},
  \bibinfo {author} {\bibfnamefont {C.}~\bibnamefont {Jun}}, \bibinfo {author}
  {\bibfnamefont {R.}~\bibnamefont {Khazaeinezhad}}, \bibinfo {author}
  {\bibfnamefont {N.}~\bibnamefont {Lippok}}, \ and\ \bibinfo {author}
  {\bibfnamefont {B.~E.}\ \bibnamefont {Bouma}},\ }\href {\doibase
  10.1364/OE.25.008255} {\bibfield  {journal} {\bibinfo  {journal} {Optics
  Express}\ }\textbf {\bibinfo {volume} {25}},\ \bibinfo {pages} {8255}
  (\bibinfo {year} {2017})}\BibitemShut {NoStop}%
\bibitem [{\citenamefont {Vakoc}\ \emph {et~al.}(2009)\citenamefont {Vakoc},
  \citenamefont {Lanning}, \citenamefont {Tyrrell}, \citenamefont {Padera},
  \citenamefont {Bartlett}, \citenamefont {Stylianopoulos}, \citenamefont
  {Munn}, \citenamefont {Tearney}, \citenamefont {Fukumura}, \citenamefont
  {Jain},\ and\ \citenamefont {Bouma}}]{Vakoc2009}%
  \BibitemOpen
  \bibfield  {author} {\bibinfo {author} {\bibfnamefont {B.~J.}\ \bibnamefont
  {Vakoc}}, \bibinfo {author} {\bibfnamefont {R.~M.}\ \bibnamefont {Lanning}},
  \bibinfo {author} {\bibfnamefont {J.~A.}\ \bibnamefont {Tyrrell}}, \bibinfo
  {author} {\bibfnamefont {T.~P.}\ \bibnamefont {Padera}}, \bibinfo {author}
  {\bibfnamefont {L.~A.}\ \bibnamefont {Bartlett}}, \bibinfo {author}
  {\bibfnamefont {T.}~\bibnamefont {Stylianopoulos}}, \bibinfo {author}
  {\bibfnamefont {L.~L.}\ \bibnamefont {Munn}}, \bibinfo {author}
  {\bibfnamefont {G.~J.}\ \bibnamefont {Tearney}}, \bibinfo {author}
  {\bibfnamefont {D.}~\bibnamefont {Fukumura}}, \bibinfo {author}
  {\bibfnamefont {R.~K.}\ \bibnamefont {Jain}}, \ and\ \bibinfo {author}
  {\bibfnamefont {B.~E.}\ \bibnamefont {Bouma}},\ }\href {\doibase
  10.1038/nm.1971} {\bibfield  {journal} {\bibinfo  {journal} {Nat. Med.}\
  }\textbf {\bibinfo {volume} {15}},\ \bibinfo {pages} {1219} (\bibinfo {year}
  {2009})}\BibitemShut {NoStop}%
\bibitem [{\citenamefont {Bolmont}\ \emph {et~al.}(2012)\citenamefont
  {Bolmont}, \citenamefont {Bouwens}, \citenamefont {Pache}, \citenamefont
  {Dimitrov}, \citenamefont {Berclaz}, \citenamefont {Villiger}, \citenamefont
  {Wegenast-Braun}, \citenamefont {Lasser},\ and\ \citenamefont
  {Fraering}}]{Bolmont2012a}%
  \BibitemOpen
  \bibfield  {author} {\bibinfo {author} {\bibfnamefont {T.}~\bibnamefont
  {Bolmont}}, \bibinfo {author} {\bibfnamefont {A.}~\bibnamefont {Bouwens}},
  \bibinfo {author} {\bibfnamefont {C.}~\bibnamefont {Pache}}, \bibinfo
  {author} {\bibfnamefont {M.}~\bibnamefont {Dimitrov}}, \bibinfo {author}
  {\bibfnamefont {C.}~\bibnamefont {Berclaz}}, \bibinfo {author} {\bibfnamefont
  {M.}~\bibnamefont {Villiger}}, \bibinfo {author} {\bibfnamefont {B.~M.}\
  \bibnamefont {Wegenast-Braun}}, \bibinfo {author} {\bibfnamefont
  {T.}~\bibnamefont {Lasser}}, \ and\ \bibinfo {author} {\bibfnamefont {P.~C.}\
  \bibnamefont {Fraering}},\ }\href {\doibase 10.1523/JNEUROSCI.0925-12.2012}
  {\bibfield  {journal} {\bibinfo  {journal} {J. Neurosci.}\ }\textbf {\bibinfo
  {volume} {32}},\ \bibinfo {pages} {14548} (\bibinfo {year}
  {2012})}\BibitemShut {NoStop}%
\bibitem [{\citenamefont {Srinivasan}\ \emph {et~al.}(2012)\citenamefont
  {Srinivasan}, \citenamefont {Radhakrishnan}, \citenamefont {Jiang},
  \citenamefont {Barry},\ and\ \citenamefont {Cable}}]{Srinivasan2012a}%
  \BibitemOpen
  \bibfield  {author} {\bibinfo {author} {\bibfnamefont {V.~J.}\ \bibnamefont
  {Srinivasan}}, \bibinfo {author} {\bibfnamefont {H.}~\bibnamefont
  {Radhakrishnan}}, \bibinfo {author} {\bibfnamefont {J.~Y.}\ \bibnamefont
  {Jiang}}, \bibinfo {author} {\bibfnamefont {S.}~\bibnamefont {Barry}}, \ and\
  \bibinfo {author} {\bibfnamefont {A.~E.}\ \bibnamefont {Cable}},\ }\href
  {\doibase 10.1364/OE.20.002220} {\bibfield  {journal} {\bibinfo  {journal}
  {Opt. Express}\ }\textbf {\bibinfo {volume} {20}},\ \bibinfo {pages} {2220}
  (\bibinfo {year} {2012})}\BibitemShut {NoStop}%
\bibitem [{\citenamefont {Leitgeb}\ \emph {et~al.}(2003)\citenamefont
  {Leitgeb}, \citenamefont {Hitzenberger},\ and\ \citenamefont
  {Fercher}}]{leitgeb_performance_2003}%
  \BibitemOpen
  \bibfield  {author} {\bibinfo {author} {\bibfnamefont {R.}~\bibnamefont
  {Leitgeb}}, \bibinfo {author} {\bibfnamefont {C.~K.}\ \bibnamefont
  {Hitzenberger}}, \ and\ \bibinfo {author} {\bibfnamefont {A.~F.}\
  \bibnamefont {Fercher}},\ }\href {\doibase 10.1364/OE.11.000889} {\bibfield
  {journal} {\bibinfo  {journal} {Opt. Express, OE}\ }\textbf {\bibinfo
  {volume} {11}},\ \bibinfo {pages} {889} (\bibinfo {year} {2003})}\BibitemShut
  {NoStop}%
\bibitem [{\citenamefont {de~Boer}\ \emph {et~al.}(2003)\citenamefont
  {de~Boer}, \citenamefont {Cense}, \citenamefont {Park}, \citenamefont
  {Pierce}, \citenamefont {Tearney},\ and\ \citenamefont {Bouma}}]{deBoer:03}%
  \BibitemOpen
  \bibfield  {author} {\bibinfo {author} {\bibfnamefont {J.~F.}\ \bibnamefont
  {de~Boer}}, \bibinfo {author} {\bibfnamefont {B.}~\bibnamefont {Cense}},
  \bibinfo {author} {\bibfnamefont {B.~H.}\ \bibnamefont {Park}}, \bibinfo
  {author} {\bibfnamefont {M.~C.}\ \bibnamefont {Pierce}}, \bibinfo {author}
  {\bibfnamefont {G.~J.}\ \bibnamefont {Tearney}}, \ and\ \bibinfo {author}
  {\bibfnamefont {B.~E.}\ \bibnamefont {Bouma}},\ }\href {\doibase
  10.1364/OL.28.002067} {\bibfield  {journal} {\bibinfo  {journal} {Opt.
  Lett.}\ }\textbf {\bibinfo {volume} {28}},\ \bibinfo {pages} {2067} (\bibinfo
  {year} {2003})}\BibitemShut {NoStop}%
\bibitem [{\citenamefont {Choma}\ \emph {et~al.}(2003)\citenamefont {Choma},
  \citenamefont {Sarunic}, \citenamefont {Yang},\ and\ \citenamefont
  {Izatt}}]{Choma:03}%
  \BibitemOpen
  \bibfield  {author} {\bibinfo {author} {\bibfnamefont {M.~A.}\ \bibnamefont
  {Choma}}, \bibinfo {author} {\bibfnamefont {M.~V.}\ \bibnamefont {Sarunic}},
  \bibinfo {author} {\bibfnamefont {C.}~\bibnamefont {Yang}}, \ and\ \bibinfo
  {author} {\bibfnamefont {J.~A.}\ \bibnamefont {Izatt}},\ }\href {\doibase
  10.1364/OE.11.002183} {\bibfield  {journal} {\bibinfo  {journal} {Opt.
  Express}\ }\textbf {\bibinfo {volume} {11}},\ \bibinfo {pages} {2183}
  (\bibinfo {year} {2003})}\BibitemShut {NoStop}%
\bibitem [{\citenamefont {Choi}\ and\ \citenamefont {Wang}(2015)}]{Choi}%
  \BibitemOpen
  \bibfield  {author} {\bibinfo {author} {\bibfnamefont {W.~J.}\ \bibnamefont
  {Choi}}\ and\ \bibinfo {author} {\bibfnamefont {R.~K.}\ \bibnamefont
  {Wang}},\ }\href {\doibase 10.1117/1.JBO.20.10} {\bibfield  {journal}
  {\bibinfo  {journal} {J. Biomed. Opt.}\ }\textbf {\bibinfo {volume} {20}},\
  \bibinfo {pages} {106004} (\bibinfo {year} {2015})}\BibitemShut {NoStop}%
\bibitem [{\citenamefont {Drexler}(2004)}]{Drexler2004}%
  \BibitemOpen
  \bibfield  {author} {\bibinfo {author} {\bibfnamefont {W.}~\bibnamefont
  {Drexler}},\ }\href {\doibase 10.1117/1.1629679} {\bibfield  {journal}
  {\bibinfo  {journal} {Journal of Biomedical Optics}\ }\textbf {\bibinfo
  {volume} {9}},\ \bibinfo {pages} {47} (\bibinfo {year} {2004})}\BibitemShut
  {NoStop}%
\bibitem [{\citenamefont {Tsai}\ \emph {et~al.}(2009)\citenamefont {Tsai},
  \citenamefont {Zhou}, \citenamefont {Adler},\ and\ \citenamefont
  {Fujimoto}}]{Tsai2009}%
  \BibitemOpen
  \bibfield  {author} {\bibinfo {author} {\bibfnamefont {T.-H.}\ \bibnamefont
  {Tsai}}, \bibinfo {author} {\bibfnamefont {C.}~\bibnamefont {Zhou}}, \bibinfo
  {author} {\bibfnamefont {D.~C.}\ \bibnamefont {Adler}}, \ and\ \bibinfo
  {author} {\bibfnamefont {J.~G.}\ \bibnamefont {Fujimoto}},\ }\href {\doibase
  10.1364/OE.17.021257} {\bibfield  {journal} {\bibinfo  {journal} {Optics
  Express}\ }\textbf {\bibinfo {volume} {17}},\ \bibinfo {pages} {21257}
  (\bibinfo {year} {2009})}\BibitemShut {NoStop}%
\bibitem [{\citenamefont {Bajraszewski}\ \emph {et~al.}(2008)\citenamefont
  {Bajraszewski}, \citenamefont {Wojtkowski}, \citenamefont {Szkulmowski},
  \citenamefont {Szkulmowska}, \citenamefont {Huber},\ and\ \citenamefont
  {Kowalczyk}}]{Bajraszewski2008a}%
  \BibitemOpen
  \bibfield  {author} {\bibinfo {author} {\bibfnamefont {T.}~\bibnamefont
  {Bajraszewski}}, \bibinfo {author} {\bibfnamefont {M.}~\bibnamefont
  {Wojtkowski}}, \bibinfo {author} {\bibfnamefont {M.}~\bibnamefont
  {Szkulmowski}}, \bibinfo {author} {\bibfnamefont {A.}~\bibnamefont
  {Szkulmowska}}, \bibinfo {author} {\bibfnamefont {R.}~\bibnamefont {Huber}},
  \ and\ \bibinfo {author} {\bibfnamefont {A.}~\bibnamefont {Kowalczyk}},\
  }\href {\doibase 10.1364/OE.16.004163} {\bibfield  {journal} {\bibinfo
  {journal} {Optics Express}\ }\textbf {\bibinfo {volume} {16}},\ \bibinfo
  {pages} {4163} (\bibinfo {year} {2008})}\BibitemShut {NoStop}%
\bibitem [{\citenamefont {Jung}\ \emph {et~al.}(2008)\citenamefont {Jung},
  \citenamefont {Park}, \citenamefont {Jeong}, \citenamefont {Kim},
  \citenamefont {Eom}, \citenamefont {Yu}, \citenamefont {Gee}, \citenamefont
  {Lee},\ and\ \citenamefont {Kim}}]{Jung:08}%
  \BibitemOpen
  \bibfield  {author} {\bibinfo {author} {\bibfnamefont {E.~J.}\ \bibnamefont
  {Jung}}, \bibinfo {author} {\bibfnamefont {J.-S.}\ \bibnamefont {Park}},
  \bibinfo {author} {\bibfnamefont {M.~Y.}\ \bibnamefont {Jeong}}, \bibinfo
  {author} {\bibfnamefont {C.-S.}\ \bibnamefont {Kim}}, \bibinfo {author}
  {\bibfnamefont {T.~J.}\ \bibnamefont {Eom}}, \bibinfo {author} {\bibfnamefont
  {B.-A.}\ \bibnamefont {Yu}}, \bibinfo {author} {\bibfnamefont
  {S.}~\bibnamefont {Gee}}, \bibinfo {author} {\bibfnamefont {J.}~\bibnamefont
  {Lee}}, \ and\ \bibinfo {author} {\bibfnamefont {M.~K.}\ \bibnamefont
  {Kim}},\ }\href {\doibase 10.1364/OE.16.017457} {\bibfield  {journal}
  {\bibinfo  {journal} {Opt. Express}\ }\textbf {\bibinfo {volume} {16}},\
  \bibinfo {pages} {17457} (\bibinfo {year} {2008})}\BibitemShut {NoStop}%
\bibitem [{\citenamefont {Del'Haye}\ \emph {et~al.}(2007)\citenamefont
  {Del'Haye}, \citenamefont {Schliesser}, \citenamefont {Arcizet},
  \citenamefont {Wilken}, \citenamefont {Holzwarth},\ and\ \citenamefont
  {Kippenberg}}]{DelHaye2007}%
  \BibitemOpen
  \bibfield  {author} {\bibinfo {author} {\bibfnamefont {P.}~\bibnamefont
  {Del'Haye}}, \bibinfo {author} {\bibfnamefont {A.}~\bibnamefont
  {Schliesser}}, \bibinfo {author} {\bibfnamefont {O.}~\bibnamefont {Arcizet}},
  \bibinfo {author} {\bibfnamefont {T.}~\bibnamefont {Wilken}}, \bibinfo
  {author} {\bibfnamefont {R.}~\bibnamefont {Holzwarth}}, \ and\ \bibinfo
  {author} {\bibfnamefont {T.~J.}\ \bibnamefont {Kippenberg}},\ }\href
  {\doibase 10.1038/nature06401} {\bibfield  {journal} {\bibinfo  {journal}
  {Nature}\ }\textbf {\bibinfo {volume} {450}},\ \bibinfo {pages} {1214}
  (\bibinfo {year} {2007})}\BibitemShut {NoStop}%
\bibitem [{\citenamefont {Kippenberg}\ \emph {et~al.}(2011)\citenamefont
  {Kippenberg}, \citenamefont {Holzwarth},\ and\ \citenamefont
  {Diddams}}]{Kippenberg2011}%
  \BibitemOpen
  \bibfield  {author} {\bibinfo {author} {\bibfnamefont {T.~J.}\ \bibnamefont
  {Kippenberg}}, \bibinfo {author} {\bibfnamefont {R.}~\bibnamefont
  {Holzwarth}}, \ and\ \bibinfo {author} {\bibfnamefont {S.~A.}\ \bibnamefont
  {Diddams}},\ }\href {\doibase 10.1126/science.1193968} {\bibfield  {journal}
  {\bibinfo  {journal} {Science}\ }\textbf {\bibinfo {volume} {332}},\ \bibinfo
  {pages} {555} (\bibinfo {year} {2011})}\BibitemShut {NoStop}%
\bibitem [{\citenamefont {Herr}\ \emph {et~al.}(2014)\citenamefont {Herr},
  \citenamefont {Brasch}, \citenamefont {Jost}, \citenamefont {Wang},
  \citenamefont {Kondratiev}, \citenamefont {Gorodetsky},\ and\ \citenamefont
  {Kippenberg}}]{Herr2013a}%
  \BibitemOpen
  \bibfield  {author} {\bibinfo {author} {\bibfnamefont {T.}~\bibnamefont
  {Herr}}, \bibinfo {author} {\bibfnamefont {V.}~\bibnamefont {Brasch}},
  \bibinfo {author} {\bibfnamefont {J.~D.}\ \bibnamefont {Jost}}, \bibinfo
  {author} {\bibfnamefont {C.~Y.}\ \bibnamefont {Wang}}, \bibinfo {author}
  {\bibfnamefont {N.~M.}\ \bibnamefont {Kondratiev}}, \bibinfo {author}
  {\bibfnamefont {M.~L.}\ \bibnamefont {Gorodetsky}}, \ and\ \bibinfo {author}
  {\bibfnamefont {T.~J.}\ \bibnamefont {Kippenberg}},\ }\href {\doibase
  10.1109/CLEOE-IQEC.2013.6801769} {\bibfield  {journal} {\bibinfo  {journal}
  {Nature Photonics}\ }\textbf {\bibinfo {volume} {8}},\ \bibinfo {pages} {145}
  (\bibinfo {year} {2014})}\BibitemShut {NoStop}%
\bibitem [{\citenamefont {Okawachi}\ \emph {et~al.}(2014)\citenamefont
  {Okawachi}, \citenamefont {Lamont}, \citenamefont {Luke}, \citenamefont
  {Carvalho}, \citenamefont {Yu}, \citenamefont {Lipson},\ and\ \citenamefont
  {Gaeta}}]{Okawachi2014}%
  \BibitemOpen
  \bibfield  {author} {\bibinfo {author} {\bibfnamefont {Y.}~\bibnamefont
  {Okawachi}}, \bibinfo {author} {\bibfnamefont {M.~R.~E.}\ \bibnamefont
  {Lamont}}, \bibinfo {author} {\bibfnamefont {K.}~\bibnamefont {Luke}},
  \bibinfo {author} {\bibfnamefont {D.~O.}\ \bibnamefont {Carvalho}}, \bibinfo
  {author} {\bibfnamefont {M.}~\bibnamefont {Yu}}, \bibinfo {author}
  {\bibfnamefont {M.}~\bibnamefont {Lipson}}, \ and\ \bibinfo {author}
  {\bibfnamefont {A.~L.}\ \bibnamefont {Gaeta}},\ }\href {\doibase
  10.1364/OL.39.003535} {\bibfield  {journal} {\bibinfo  {journal} {Optics
  Letters}\ }\textbf {\bibinfo {volume} {39}},\ \bibinfo {pages} {3535}
  (\bibinfo {year} {2014})}\BibitemShut {NoStop}%
\bibitem [{\citenamefont {Pfeiffer}\ \emph {et~al.}(2017)\citenamefont
  {Pfeiffer}, \citenamefont {Herkommer}, \citenamefont {Liu}, \citenamefont
  {Guo}, \citenamefont {Karpov}, \citenamefont {Lucas}, \citenamefont
  {Zervas},\ and\ \citenamefont {Kippenberg}}]{Pfeiffer2017}%
  \BibitemOpen
  \bibfield  {author} {\bibinfo {author} {\bibfnamefont {M.~H.~P.}\
  \bibnamefont {Pfeiffer}}, \bibinfo {author} {\bibfnamefont {C.}~\bibnamefont
  {Herkommer}}, \bibinfo {author} {\bibfnamefont {J.}~\bibnamefont {Liu}},
  \bibinfo {author} {\bibfnamefont {H.}~\bibnamefont {Guo}}, \bibinfo {author}
  {\bibfnamefont {M.}~\bibnamefont {Karpov}}, \bibinfo {author} {\bibfnamefont
  {E.}~\bibnamefont {Lucas}}, \bibinfo {author} {\bibfnamefont
  {M.}~\bibnamefont {Zervas}}, \ and\ \bibinfo {author} {\bibfnamefont {T.~J.}\
  \bibnamefont {Kippenberg}},\ }\href {\doibase 10.1364/OPTICA.4.000684}
  {\bibfield  {journal} {\bibinfo  {journal} {Optica}\ }\textbf {\bibinfo
  {volume} {4}},\ \bibinfo {pages} {684} (\bibinfo {year} {2017})}\BibitemShut
  {NoStop}%
\bibitem [{\citenamefont {Liu}\ \emph {et~al.}(2018{\natexlab{a}})\citenamefont
  {Liu}, \citenamefont {Raja}, \citenamefont {Karpov}, \citenamefont
  {Ghadiani}, \citenamefont {Pfeiffer}, \citenamefont {Du}, \citenamefont
  {Engelsen}, \citenamefont {Guo}, \citenamefont {Zervas},\ and\ \citenamefont
  {Kippenberg}}]{Liu:18a}%
  \BibitemOpen
  \bibfield  {author} {\bibinfo {author} {\bibfnamefont {J.}~\bibnamefont
  {Liu}}, \bibinfo {author} {\bibfnamefont {A.~S.}\ \bibnamefont {Raja}},
  \bibinfo {author} {\bibfnamefont {M.}~\bibnamefont {Karpov}}, \bibinfo
  {author} {\bibfnamefont {B.}~\bibnamefont {Ghadiani}}, \bibinfo {author}
  {\bibfnamefont {M.~H.~P.}\ \bibnamefont {Pfeiffer}}, \bibinfo {author}
  {\bibfnamefont {B.}~\bibnamefont {Du}}, \bibinfo {author} {\bibfnamefont
  {N.~J.}\ \bibnamefont {Engelsen}}, \bibinfo {author} {\bibfnamefont
  {H.}~\bibnamefont {Guo}}, \bibinfo {author} {\bibfnamefont {M.}~\bibnamefont
  {Zervas}}, \ and\ \bibinfo {author} {\bibfnamefont {T.~J.}\ \bibnamefont
  {Kippenberg}},\ }\href {\doibase 10.1364/OPTICA.5.001347} {\bibfield
  {journal} {\bibinfo  {journal} {Optica}\ }\textbf {\bibinfo {volume} {5}},\
  \bibinfo {pages} {1347} (\bibinfo {year} {2018}{\natexlab{a}})}\BibitemShut
  {NoStop}%
\bibitem [{\citenamefont {Stern}\ \emph {et~al.}(2018)\citenamefont {Stern},
  \citenamefont {Ji}, \citenamefont {Okawachi}, \citenamefont {Gaeta},\ and\
  \citenamefont {Lipson}}]{Stern:2018ca}%
  \BibitemOpen
  \bibfield  {author} {\bibinfo {author} {\bibfnamefont {B.}~\bibnamefont
  {Stern}}, \bibinfo {author} {\bibfnamefont {X.}~\bibnamefont {Ji}}, \bibinfo
  {author} {\bibfnamefont {Y.}~\bibnamefont {Okawachi}}, \bibinfo {author}
  {\bibfnamefont {A.~L.}\ \bibnamefont {Gaeta}}, \ and\ \bibinfo {author}
  {\bibfnamefont {M.}~\bibnamefont {Lipson}},\ }\href {\doibase
  10.1038/s41586-018-0598-9} {\bibfield  {journal} {\bibinfo  {journal}
  {Nature}\ }\textbf {\bibinfo {volume} {562}},\ \bibinfo {pages} {401}
  (\bibinfo {year} {2018})}\BibitemShut {NoStop}%
\bibitem [{\citenamefont {Pfeiffer}\ \emph {et~al.}(2016)\citenamefont
  {Pfeiffer}, \citenamefont {Kordts}, \citenamefont {Brasch}, \citenamefont
  {Zervas}, \citenamefont {Geiselmann}, \citenamefont {Jost},\ and\
  \citenamefont {Kippenberg}}]{Pfeiffer2016}%
  \BibitemOpen
  \bibfield  {author} {\bibinfo {author} {\bibfnamefont {M.~H.~P.}\
  \bibnamefont {Pfeiffer}}, \bibinfo {author} {\bibfnamefont {A.}~\bibnamefont
  {Kordts}}, \bibinfo {author} {\bibfnamefont {V.}~\bibnamefont {Brasch}},
  \bibinfo {author} {\bibfnamefont {M.}~\bibnamefont {Zervas}}, \bibinfo
  {author} {\bibfnamefont {M.}~\bibnamefont {Geiselmann}}, \bibinfo {author}
  {\bibfnamefont {J.~D.}\ \bibnamefont {Jost}}, \ and\ \bibinfo {author}
  {\bibfnamefont {T.~J.}\ \bibnamefont {Kippenberg}},\ }\href {\doibase
  10.1364/OPTICA.3.000020} {\bibfield  {journal} {\bibinfo  {journal} {Optica}\
  }\textbf {\bibinfo {volume} {3}},\ \bibinfo {pages} {20} (\bibinfo {year}
  {2016})}\BibitemShut {NoStop}%
\bibitem [{\citenamefont {Guo}\ \emph {et~al.}(2016)\citenamefont {Guo},
  \citenamefont {Karpov}, \citenamefont {Lucas}, \citenamefont {Kordts},
  \citenamefont {Pfeiffer}, \citenamefont {Brasch}, \citenamefont {Lihachev},
  \citenamefont {Lobanov}, \citenamefont {Gorodetsky},\ and\ \citenamefont
  {Kippenberg}}]{Guo2016}%
  \BibitemOpen
  \bibfield  {author} {\bibinfo {author} {\bibfnamefont {H.}~\bibnamefont
  {Guo}}, \bibinfo {author} {\bibfnamefont {M.}~\bibnamefont {Karpov}},
  \bibinfo {author} {\bibfnamefont {E.}~\bibnamefont {Lucas}}, \bibinfo
  {author} {\bibfnamefont {A.}~\bibnamefont {Kordts}}, \bibinfo {author}
  {\bibfnamefont {M.~H.~P.}\ \bibnamefont {Pfeiffer}}, \bibinfo {author}
  {\bibfnamefont {V.}~\bibnamefont {Brasch}}, \bibinfo {author} {\bibfnamefont
  {G.}~\bibnamefont {Lihachev}}, \bibinfo {author} {\bibfnamefont {V.~E.}\
  \bibnamefont {Lobanov}}, \bibinfo {author} {\bibfnamefont {M.~L.}\
  \bibnamefont {Gorodetsky}}, \ and\ \bibinfo {author} {\bibfnamefont {T.~J.}\
  \bibnamefont {Kippenberg}},\ }\href {\doibase 10.1038/nphys3893} {\bibfield
  {journal} {\bibinfo  {journal} {Nature Physics}\ }\textbf {\bibinfo {volume}
  {13}},\ \bibinfo {pages} {94} (\bibinfo {year} {2016})}\BibitemShut {NoStop}%
\bibitem [{\citenamefont {Herr}\ \emph {et~al.}(2012)\citenamefont {Herr},
  \citenamefont {Hartinger}, \citenamefont {Riemensberger}, \citenamefont
  {Wang}, \citenamefont {Gavartin}, \citenamefont {Holzwarth}, \citenamefont
  {Gorodetsky},\ and\ \citenamefont {Kippenberg}}]{herr_universal_2012}%
  \BibitemOpen
  \bibfield  {author} {\bibinfo {author} {\bibfnamefont {T.}~\bibnamefont
  {Herr}}, \bibinfo {author} {\bibfnamefont {K.}~\bibnamefont {Hartinger}},
  \bibinfo {author} {\bibfnamefont {J.}~\bibnamefont {Riemensberger}}, \bibinfo
  {author} {\bibfnamefont {C.~Y.}\ \bibnamefont {Wang}}, \bibinfo {author}
  {\bibfnamefont {E.}~\bibnamefont {Gavartin}}, \bibinfo {author}
  {\bibfnamefont {R.}~\bibnamefont {Holzwarth}}, \bibinfo {author}
  {\bibfnamefont {M.~L.}\ \bibnamefont {Gorodetsky}}, \ and\ \bibinfo {author}
  {\bibfnamefont {T.~J.}\ \bibnamefont {Kippenberg}},\ }\href
  {https://doi.org/10.1038/nphoton.2012.127} {\bibfield  {journal} {\bibinfo
  {journal} {Nature Photonics}\ }\textbf {\bibinfo {volume} {6}},\ \bibinfo
  {pages} {480} (\bibinfo {year} {2012})}\BibitemShut {NoStop}%
\bibitem [{\citenamefont {Yurek}\ \emph {et~al.}(1986)\citenamefont {Yurek},
  \citenamefont {Taylor}, \citenamefont {Goldberg}, \citenamefont {Weller},\
  and\ \citenamefont {Dandridge}}]{yurek1986quantum}%
  \BibitemOpen
  \bibfield  {author} {\bibinfo {author} {\bibfnamefont {A.}~\bibnamefont
  {Yurek}}, \bibinfo {author} {\bibfnamefont {H.}~\bibnamefont {Taylor}},
  \bibinfo {author} {\bibfnamefont {L.}~\bibnamefont {Goldberg}}, \bibinfo
  {author} {\bibfnamefont {J.}~\bibnamefont {Weller}}, \ and\ \bibinfo {author}
  {\bibfnamefont {A.}~\bibnamefont {Dandridge}},\ }\href@noop {} {\bibfield
  {journal} {\bibinfo  {journal} {IEEE journal of quantum electronics}\
  }\textbf {\bibinfo {volume} {22}},\ \bibinfo {pages} {522} (\bibinfo {year}
  {1986})}\BibitemShut {NoStop}%
\bibitem [{\citenamefont {Baney}(1998)}]{dericksen13}%
  \BibitemOpen
  \bibfield  {author} {\bibinfo {author} {\bibfnamefont {D.~M.}\ \bibnamefont
  {Baney}},\ }in\ \href@noop {} {\emph {\bibinfo {booktitle} {Fiber-optic test
  and measurement}}},\ \bibinfo {editor} {edited by\ \bibinfo {editor}
  {\bibfnamefont {D.}~\bibnamefont {Dericksen}}}\ (\bibinfo  {publisher}
  {Prentice-Hall},\ \bibinfo {address} {Englewood Cliffs, NJ},\ \bibinfo {year}
  {1998})\ Chap.~\bibinfo {chapter} {13}\BibitemShut {NoStop}%
\bibitem [{\citenamefont {Sorin}\ and\ \citenamefont
  {Baney}(1992)}]{sorin1992simple}%
  \BibitemOpen
  \bibfield  {author} {\bibinfo {author} {\bibfnamefont {W.~V.}\ \bibnamefont
  {Sorin}}\ and\ \bibinfo {author} {\bibfnamefont {D.~M.}\ \bibnamefont
  {Baney}},\ }\href@noop {} {\bibfield  {journal} {\bibinfo  {journal} {IEEE
  Photonics Technology Letters}\ }\textbf {\bibinfo {volume} {4}},\ \bibinfo
  {pages} {1404} (\bibinfo {year} {1992})}\BibitemShut {NoStop}%
\bibitem [{\citenamefont {Yun}\ \emph {et~al.}(2004)\citenamefont {Yun},
  \citenamefont {Tearney}, \citenamefont {de~Boer},\ and\ \citenamefont
  {Bouma}}]{yun2004pulsed}%
  \BibitemOpen
  \bibfield  {author} {\bibinfo {author} {\bibfnamefont {S.}~\bibnamefont
  {Yun}}, \bibinfo {author} {\bibfnamefont {G.}~\bibnamefont {Tearney}},
  \bibinfo {author} {\bibfnamefont {J.}~\bibnamefont {de~Boer}}, \ and\
  \bibinfo {author} {\bibfnamefont {B.}~\bibnamefont {Bouma}},\ }\href@noop {}
  {\bibfield  {journal} {\bibinfo  {journal} {Optics Express}\ }\textbf
  {\bibinfo {volume} {12}},\ \bibinfo {pages} {5614} (\bibinfo {year}
  {2004})}\BibitemShut {NoStop}%
\bibitem [{\citenamefont {Siegman}\ and\ \citenamefont
  {Siegman}(1971)}]{siegman1971introduction}%
  \BibitemOpen
  \bibfield  {author} {\bibinfo {author} {\bibfnamefont {A.~E.}\ \bibnamefont
  {Siegman}}\ and\ \bibinfo {author} {\bibfnamefont {A.}~\bibnamefont
  {Siegman}},\ }\href@noop {} {\emph {\bibinfo {title} {An introduction to
  lasers and masers}}},\ Vol.\ \bibinfo {volume} {122}\ (\bibinfo  {publisher}
  {McGraw-Hill New York},\ \bibinfo {year} {1971})\BibitemShut {NoStop}%
\bibitem [{\citenamefont {Izatt}\ and\ \citenamefont
  {Choma}(2008)}]{Izatt2008}%
  \BibitemOpen
  \bibfield  {author} {\bibinfo {author} {\bibfnamefont {J.~A.}\ \bibnamefont
  {Izatt}}\ and\ \bibinfo {author} {\bibfnamefont {M.~A.}\ \bibnamefont
  {Choma}},\ }in\ \href {\doibase 10.1007/978-3-540-77550-8_2} {\emph {\bibinfo
  {booktitle} {Optical Coherence Tomography}}}\ (\bibinfo  {publisher}
  {Springer},\ \bibinfo {year} {2008})\ pp.\ \bibinfo {pages}
  {47--72}\BibitemShut {NoStop}%
\bibitem [{\citenamefont {Lichtenegger}\ \emph {et~al.}(2018)\citenamefont
  {Lichtenegger}, \citenamefont {Eugui}, \citenamefont {Harper}, \citenamefont
  {Hitzenberger}, \citenamefont {Baumann}, \citenamefont {Lichtenegger},
  \citenamefont {Muck}, \citenamefont {Eugui}, \citenamefont {Harper},
  \citenamefont {Augustin}, \citenamefont {Leskovar}, \citenamefont
  {Hitzenberger}, \citenamefont {Woehrer},\ and\ \citenamefont
  {Baumann}}]{Lichtenegger2018}%
  \BibitemOpen
  \bibfield  {author} {\bibinfo {author} {\bibfnamefont {A.}~\bibnamefont
  {Lichtenegger}}, \bibinfo {author} {\bibfnamefont {P.}~\bibnamefont {Eugui}},
  \bibinfo {author} {\bibfnamefont {D.~J.}\ \bibnamefont {Harper}}, \bibinfo
  {author} {\bibfnamefont {C.~K.}\ \bibnamefont {Hitzenberger}}, \bibinfo
  {author} {\bibfnamefont {B.}~\bibnamefont {Baumann}}, \bibinfo {author}
  {\bibfnamefont {A.}~\bibnamefont {Lichtenegger}}, \bibinfo {author}
  {\bibfnamefont {M.}~\bibnamefont {Muck}}, \bibinfo {author} {\bibfnamefont
  {P.}~\bibnamefont {Eugui}}, \bibinfo {author} {\bibfnamefont {D.~J.}\
  \bibnamefont {Harper}}, \bibinfo {author} {\bibfnamefont {M.}~\bibnamefont
  {Augustin}}, \bibinfo {author} {\bibfnamefont {K.}~\bibnamefont {Leskovar}},
  \bibinfo {author} {\bibfnamefont {C.~K.}\ \bibnamefont {Hitzenberger}},
  \bibinfo {author} {\bibfnamefont {A.}~\bibnamefont {Woehrer}}, \ and\
  \bibinfo {author} {\bibfnamefont {B.}~\bibnamefont {Baumann}},\ }\href
  {\doibase 10.1117/1.NPh.5.3.035002} {\bibfield  {journal} {\bibinfo
  {journal} {Neurophotonics}\ }\textbf {\bibinfo {volume} {5}},\ \bibinfo
  {pages} {5 } (\bibinfo {year} {2018})}\BibitemShut {NoStop}%
\bibitem [{\citenamefont {Assayag}\ \emph {et~al.}(2013)\citenamefont
  {Assayag}, \citenamefont {Grieve}, \citenamefont {Devaux}, \citenamefont
  {Harms}, \citenamefont {Pallud}, \citenamefont {Chretien}, \citenamefont
  {Boccara},\ and\ \citenamefont {Varlet}}]{Assayag2013}%
  \BibitemOpen
  \bibfield  {author} {\bibinfo {author} {\bibfnamefont {O.}~\bibnamefont
  {Assayag}}, \bibinfo {author} {\bibfnamefont {K.}~\bibnamefont {Grieve}},
  \bibinfo {author} {\bibfnamefont {B.}~\bibnamefont {Devaux}}, \bibinfo
  {author} {\bibfnamefont {F.}~\bibnamefont {Harms}}, \bibinfo {author}
  {\bibfnamefont {J.}~\bibnamefont {Pallud}}, \bibinfo {author} {\bibfnamefont
  {F.}~\bibnamefont {Chretien}}, \bibinfo {author} {\bibfnamefont
  {C.}~\bibnamefont {Boccara}}, \ and\ \bibinfo {author} {\bibfnamefont
  {P.}~\bibnamefont {Varlet}},\ }\href {\doibase 10.1016/j.nicl.2013.04.005}
  {\bibfield  {journal} {\bibinfo  {journal} {Neuroimage Clin.}\ }\textbf
  {\bibinfo {volume} {2}},\ \bibinfo {pages} {549} (\bibinfo {year}
  {2013})}\BibitemShut {NoStop}%
\bibitem [{\citenamefont {Villiger}\ \emph {et~al.}(2010)\citenamefont
  {Villiger}, \citenamefont {Pache},\ and\ \citenamefont
  {Lasser}}]{Villiger:10}%
  \BibitemOpen
  \bibfield  {author} {\bibinfo {author} {\bibfnamefont {M.}~\bibnamefont
  {Villiger}}, \bibinfo {author} {\bibfnamefont {C.}~\bibnamefont {Pache}}, \
  and\ \bibinfo {author} {\bibfnamefont {T.}~\bibnamefont {Lasser}},\ }\href
  {\doibase 10.1364/OL.35.003489} {\bibfield  {journal} {\bibinfo  {journal}
  {Opt. Lett.}\ }\textbf {\bibinfo {volume} {35}},\ \bibinfo {pages} {3489}
  (\bibinfo {year} {2010})}\BibitemShut {NoStop}%
\bibitem [{\citenamefont {Blatter}\ \emph {et~al.}(2011)\citenamefont
  {Blatter}, \citenamefont {Grajciar}, \citenamefont {Eigenwillig},
  \citenamefont {Wieser}, \citenamefont {Biedermann}, \citenamefont {Huber},\
  and\ \citenamefont {Leitgeb}}]{Blatter2011}%
  \BibitemOpen
  \bibfield  {author} {\bibinfo {author} {\bibfnamefont {C.}~\bibnamefont
  {Blatter}}, \bibinfo {author} {\bibfnamefont {B.}~\bibnamefont {Grajciar}},
  \bibinfo {author} {\bibfnamefont {C.~M.}\ \bibnamefont {Eigenwillig}},
  \bibinfo {author} {\bibfnamefont {W.}~\bibnamefont {Wieser}}, \bibinfo
  {author} {\bibfnamefont {B.~R.}\ \bibnamefont {Biedermann}}, \bibinfo
  {author} {\bibfnamefont {R.}~\bibnamefont {Huber}}, \ and\ \bibinfo {author}
  {\bibfnamefont {R.~A.}\ \bibnamefont {Leitgeb}},\ }\href
  {http://www.ncbi.nlm.nih.gov/pubmed/21716451} {\bibfield  {journal} {\bibinfo
   {journal} {Opt. Express}\ }\textbf {\bibinfo {volume} {19}},\ \bibinfo
  {pages} {12141} (\bibinfo {year} {2011})}\BibitemShut {NoStop}%
\bibitem [{\citenamefont {Amirsolaimani}\ \emph {et~al.}(2017)\citenamefont
  {Amirsolaimani}, \citenamefont {Cromey}, \citenamefont {Peyghambarian},\ and\
  \citenamefont {Kieu}}]{Amirsolaimani:17}%
  \BibitemOpen
  \bibfield  {author} {\bibinfo {author} {\bibfnamefont {B.}~\bibnamefont
  {Amirsolaimani}}, \bibinfo {author} {\bibfnamefont {B.}~\bibnamefont
  {Cromey}}, \bibinfo {author} {\bibfnamefont {N.}~\bibnamefont
  {Peyghambarian}}, \ and\ \bibinfo {author} {\bibfnamefont {K.}~\bibnamefont
  {Kieu}},\ }\href {\doibase 10.1364/OE.25.023399} {\bibfield  {journal}
  {\bibinfo  {journal} {Opt. Express}\ }\textbf {\bibinfo {volume} {25}},\
  \bibinfo {pages} {23399} (\bibinfo {year} {2017})}\BibitemShut {NoStop}%
\bibitem [{\citenamefont {Belykh}\ \emph
  {et~al.}(2018{\natexlab{a}})\citenamefont {Belykh}, \citenamefont {Cavallo},
  \citenamefont {Gandhi}, \citenamefont {Zhao}, \citenamefont {Veljanoski},
  \citenamefont {Izady}, \citenamefont {Martirosyan}, \citenamefont
  {Byvaltsev}, \citenamefont {Eschbacher}, \citenamefont {Preul} \emph
  {et~al.}}]{belykh2018utilization}%
  \BibitemOpen
  \bibfield  {author} {\bibinfo {author} {\bibfnamefont {E.}~\bibnamefont
  {Belykh}}, \bibinfo {author} {\bibfnamefont {C.}~\bibnamefont {Cavallo}},
  \bibinfo {author} {\bibfnamefont {S.}~\bibnamefont {Gandhi}}, \bibinfo
  {author} {\bibfnamefont {X.}~\bibnamefont {Zhao}}, \bibinfo {author}
  {\bibfnamefont {D.}~\bibnamefont {Veljanoski}}, \bibinfo {author}
  {\bibfnamefont {M.~Y.}\ \bibnamefont {Izady}}, \bibinfo {author}
  {\bibfnamefont {N.}~\bibnamefont {Martirosyan}}, \bibinfo {author}
  {\bibfnamefont {V.}~\bibnamefont {Byvaltsev}}, \bibinfo {author}
  {\bibfnamefont {J.}~\bibnamefont {Eschbacher}}, \bibinfo {author}
  {\bibfnamefont {M.}~\bibnamefont {Preul}},  \emph {et~al.},\ }\href {\doibase
  10.23736/S0390-5616.18.04553-8} {\bibfield  {journal} {\bibinfo  {journal}
  {Journal of neurosurgical sciences}\ }\textbf {\bibinfo {volume} {62}},\
  \bibinfo {pages} {704} (\bibinfo {year} {2018}{\natexlab{a}})}\BibitemShut
  {NoStop}%
\bibitem [{\citenamefont {Sanai}\ \emph {et~al.}(2011)\citenamefont {Sanai},
  \citenamefont {Eschbacher}, \citenamefont {Hattendorf}, \citenamefont
  {Coons}, \citenamefont {Preul}, \citenamefont {Smith}, \citenamefont
  {Nakaji},\ and\ \citenamefont {Spetzler}}]{sanai2011intraoperative}%
  \BibitemOpen
  \bibfield  {author} {\bibinfo {author} {\bibfnamefont {N.}~\bibnamefont
  {Sanai}}, \bibinfo {author} {\bibfnamefont {J.}~\bibnamefont {Eschbacher}},
  \bibinfo {author} {\bibfnamefont {G.}~\bibnamefont {Hattendorf}}, \bibinfo
  {author} {\bibfnamefont {S.~W.}\ \bibnamefont {Coons}}, \bibinfo {author}
  {\bibfnamefont {M.~C.}\ \bibnamefont {Preul}}, \bibinfo {author}
  {\bibfnamefont {K.~A.}\ \bibnamefont {Smith}}, \bibinfo {author}
  {\bibfnamefont {P.}~\bibnamefont {Nakaji}}, \ and\ \bibinfo {author}
  {\bibfnamefont {R.~F.}\ \bibnamefont {Spetzler}},\ }\href {\doibase
  10.1227/NEU.0b013e318212464e} {\bibfield  {journal} {\bibinfo  {journal}
  {Operative Neurosurgery}\ }\textbf {\bibinfo {volume} {68}},\ \bibinfo
  {pages} {ons282} (\bibinfo {year} {2011})}\BibitemShut {NoStop}%
\bibitem [{\citenamefont {Schlosser}\ and\ \citenamefont
  {Bojarski}(2011)}]{schlosser2011confocal}%
  \BibitemOpen
  \bibfield  {author} {\bibinfo {author} {\bibfnamefont {H.-G.}\ \bibnamefont
  {Schlosser}}\ and\ \bibinfo {author} {\bibfnamefont {C.}~\bibnamefont
  {Bojarski}},\ }in\ \href {\doibase 10.5772/24808} {\emph {\bibinfo
  {booktitle} {Advances in the Biology, Imaging and Therapies for
  Glioblastoma}}}\ (\bibinfo  {publisher} {InTech},\ \bibinfo {year}
  {2011})\BibitemShut {NoStop}%
\bibitem [{\citenamefont {Charalampaki}\ \emph {et~al.}(2015)\citenamefont
  {Charalampaki}, \citenamefont {Javed}, \citenamefont {Daali}, \citenamefont
  {Heiroth}, \citenamefont {Igressa},\ and\ \citenamefont
  {Weber}}]{charalampaki2015confocal}%
  \BibitemOpen
  \bibfield  {author} {\bibinfo {author} {\bibfnamefont {P.}~\bibnamefont
  {Charalampaki}}, \bibinfo {author} {\bibfnamefont {M.}~\bibnamefont {Javed}},
  \bibinfo {author} {\bibfnamefont {S.}~\bibnamefont {Daali}}, \bibinfo
  {author} {\bibfnamefont {H.-J.}\ \bibnamefont {Heiroth}}, \bibinfo {author}
  {\bibfnamefont {A.}~\bibnamefont {Igressa}}, \ and\ \bibinfo {author}
  {\bibfnamefont {F.}~\bibnamefont {Weber}},\ }\href {\doibase 10.5772/24808}
  {\bibfield  {journal} {\bibinfo  {journal} {Neurosurgery}\ }\textbf {\bibinfo
  {volume} {62}},\ \bibinfo {pages} {171} (\bibinfo {year} {2015})}\BibitemShut
  {NoStop}%
\bibitem [{\citenamefont {Zehri}\ \emph {et~al.}(2014)\citenamefont {Zehri},
  \citenamefont {Wyatt~Ramey}, \citenamefont {Mooney}, \citenamefont
  {Martirosyan}, \citenamefont {Preul},\ and\ \citenamefont
  {Nakaji}}]{zehri2014neurosurgical}%
  \BibitemOpen
  \bibfield  {author} {\bibinfo {author} {\bibfnamefont {A.~H.}\ \bibnamefont
  {Zehri}}, \bibinfo {author} {\bibfnamefont {J.~F.~G.}\ \bibnamefont
  {Wyatt~Ramey}}, \bibinfo {author} {\bibfnamefont {M.~A.}\ \bibnamefont
  {Mooney}}, \bibinfo {author} {\bibfnamefont {N.~L.}\ \bibnamefont
  {Martirosyan}}, \bibinfo {author} {\bibfnamefont {M.~C.}\ \bibnamefont
  {Preul}}, \ and\ \bibinfo {author} {\bibfnamefont {P.}~\bibnamefont
  {Nakaji}},\ }\href {\doibase 10.4103/2152-7806.131638} {\bibfield  {journal}
  {\bibinfo  {journal} {Surgical neurology international}\ }\textbf {\bibinfo
  {volume} {5}},\ \bibinfo {pages} {60} (\bibinfo {year} {2014})}\BibitemShut
  {NoStop}%
\bibitem [{\citenamefont {Fugazza}\ \emph {et~al.}(2016)\citenamefont
  {Fugazza}, \citenamefont {Gaiani}, \citenamefont {Carra}, \citenamefont
  {Brunetti}, \citenamefont {L{\'e}vy}, \citenamefont {Sobhani}, \citenamefont
  {Azoulay}, \citenamefont {Catena}, \citenamefont {de'Angelis},\ and\
  \citenamefont {de'Angelis}}]{fugazza2016confocal}%
  \BibitemOpen
  \bibfield  {author} {\bibinfo {author} {\bibfnamefont {A.}~\bibnamefont
  {Fugazza}}, \bibinfo {author} {\bibfnamefont {F.}~\bibnamefont {Gaiani}},
  \bibinfo {author} {\bibfnamefont {M.~C.}\ \bibnamefont {Carra}}, \bibinfo
  {author} {\bibfnamefont {F.}~\bibnamefont {Brunetti}}, \bibinfo {author}
  {\bibfnamefont {M.}~\bibnamefont {L{\'e}vy}}, \bibinfo {author}
  {\bibfnamefont {I.}~\bibnamefont {Sobhani}}, \bibinfo {author} {\bibfnamefont
  {D.}~\bibnamefont {Azoulay}}, \bibinfo {author} {\bibfnamefont
  {F.}~\bibnamefont {Catena}}, \bibinfo {author} {\bibfnamefont {G.~L.}\
  \bibnamefont {de'Angelis}}, \ and\ \bibinfo {author} {\bibfnamefont
  {N.}~\bibnamefont {de'Angelis}},\ }\href {\doibase 10.1155/2016/4638683}
  {\bibfield  {journal} {\bibinfo  {journal} {BioMed research international}\
  }\textbf {\bibinfo {volume} {2016}},\ \bibinfo {pages} {31} (\bibinfo {year}
  {2016})}\BibitemShut {NoStop}%
\bibitem [{\citenamefont {Wellikoff}\ \emph {et~al.}(2015)\citenamefont
  {Wellikoff}, \citenamefont {Holladay}, \citenamefont {Downie}, \citenamefont
  {Chaudoir}, \citenamefont {Brandi},\ and\ \citenamefont
  {Turbat-Herrera}}]{wellikoff2015comparison}%
  \BibitemOpen
  \bibfield  {author} {\bibinfo {author} {\bibfnamefont {A.~S.}\ \bibnamefont
  {Wellikoff}}, \bibinfo {author} {\bibfnamefont {R.~C.}\ \bibnamefont
  {Holladay}}, \bibinfo {author} {\bibfnamefont {G.~H.}\ \bibnamefont
  {Downie}}, \bibinfo {author} {\bibfnamefont {C.~S.}\ \bibnamefont
  {Chaudoir}}, \bibinfo {author} {\bibfnamefont {L.}~\bibnamefont {Brandi}}, \
  and\ \bibinfo {author} {\bibfnamefont {E.~A.}\ \bibnamefont
  {Turbat-Herrera}},\ }\href {\doibase 10.1111/resp.12578} {\bibfield
  {journal} {\bibinfo  {journal} {Respirology}\ }\textbf {\bibinfo {volume}
  {20}},\ \bibinfo {pages} {967} (\bibinfo {year} {2015})}\BibitemShut
  {NoStop}%
\bibitem [{\citenamefont {Bui}\ \emph {et~al.}(2015)\citenamefont {Bui},
  \citenamefont {Mach}, \citenamefont {Zlatev}, \citenamefont {Rouse},
  \citenamefont {Leppert},\ and\ \citenamefont {Liao}}]{bui2015pilot}%
  \BibitemOpen
  \bibfield  {author} {\bibinfo {author} {\bibfnamefont {D.}~\bibnamefont
  {Bui}}, \bibinfo {author} {\bibfnamefont {K.~E.}\ \bibnamefont {Mach}},
  \bibinfo {author} {\bibfnamefont {D.~V.}\ \bibnamefont {Zlatev}}, \bibinfo
  {author} {\bibfnamefont {R.~V.}\ \bibnamefont {Rouse}}, \bibinfo {author}
  {\bibfnamefont {J.~T.}\ \bibnamefont {Leppert}}, \ and\ \bibinfo {author}
  {\bibfnamefont {J.~C.}\ \bibnamefont {Liao}},\ }\href {\doibase
  10.1089/end.2015.0523} {\bibfield  {journal} {\bibinfo  {journal} {Journal of
  endourology}\ }\textbf {\bibinfo {volume} {29}},\ \bibinfo {pages} {1418}
  (\bibinfo {year} {2015})}\BibitemShut {NoStop}%
\bibitem [{\citenamefont {Fuks}\ \emph {et~al.}(2018)\citenamefont {Fuks},
  \citenamefont {Pierangelo}, \citenamefont {Validire}, \citenamefont
  {Lefevre}, \citenamefont {Benali}, \citenamefont {Trebuchet}, \citenamefont
  {Criton},\ and\ \citenamefont {Gayet}}]{fuks2018intraoperative}%
  \BibitemOpen
  \bibfield  {author} {\bibinfo {author} {\bibfnamefont {D.}~\bibnamefont
  {Fuks}}, \bibinfo {author} {\bibfnamefont {A.}~\bibnamefont {Pierangelo}},
  \bibinfo {author} {\bibfnamefont {P.}~\bibnamefont {Validire}}, \bibinfo
  {author} {\bibfnamefont {M.}~\bibnamefont {Lefevre}}, \bibinfo {author}
  {\bibfnamefont {A.}~\bibnamefont {Benali}}, \bibinfo {author} {\bibfnamefont
  {G.}~\bibnamefont {Trebuchet}}, \bibinfo {author} {\bibfnamefont
  {A.}~\bibnamefont {Criton}}, \ and\ \bibinfo {author} {\bibfnamefont
  {B.}~\bibnamefont {Gayet}},\ }\href {\doibase 10.1007/s00464-018-6442-3}
  {\bibfield  {journal} {\bibinfo  {journal} {Surgical endoscopy}\ ,\ \bibinfo
  {pages} {1}} (\bibinfo {year} {2018})}\BibitemShut {NoStop}%
\bibitem [{\citenamefont {Krafft}\ \emph {et~al.}(2018)\citenamefont {Krafft},
  \citenamefont {von Eggeling}, \citenamefont {Guntinas-Lichius}, \citenamefont
  {Hartmann}, \citenamefont {Waldner}, \citenamefont {Neurath},\ and\
  \citenamefont {Popp}}]{krafft2018perspectives}%
  \BibitemOpen
  \bibfield  {author} {\bibinfo {author} {\bibfnamefont {C.}~\bibnamefont
  {Krafft}}, \bibinfo {author} {\bibfnamefont {F.}~\bibnamefont {von
  Eggeling}}, \bibinfo {author} {\bibfnamefont {O.}~\bibnamefont
  {Guntinas-Lichius}}, \bibinfo {author} {\bibfnamefont {A.}~\bibnamefont
  {Hartmann}}, \bibinfo {author} {\bibfnamefont {M.~J.}\ \bibnamefont
  {Waldner}}, \bibinfo {author} {\bibfnamefont {M.~F.}\ \bibnamefont
  {Neurath}}, \ and\ \bibinfo {author} {\bibfnamefont {J.}~\bibnamefont
  {Popp}},\ }\href {\doibase 10.1002/jbio.201700236} {\bibfield  {journal}
  {\bibinfo  {journal} {Journal of biophotonics}\ }\textbf {\bibinfo {volume}
  {11}},\ \bibinfo {pages} {e201700236} (\bibinfo {year} {2018})}\BibitemShut
  {NoStop}%
\bibitem [{\citenamefont {Belykh}\ \emph
  {et~al.}(2018{\natexlab{b}})\citenamefont {Belykh}, \citenamefont {Patel},
  \citenamefont {Miller}, \citenamefont {Bozkurt}, \citenamefont
  {Ya{\u{g}}murlu}, \citenamefont {Woolf}, \citenamefont {Scheck},
  \citenamefont {Eschbacher}, \citenamefont {Nakaji},\ and\ \citenamefont
  {Preul}}]{belykh2018probe}%
  \BibitemOpen
  \bibfield  {author} {\bibinfo {author} {\bibfnamefont {E.}~\bibnamefont
  {Belykh}}, \bibinfo {author} {\bibfnamefont {A.~A.}\ \bibnamefont {Patel}},
  \bibinfo {author} {\bibfnamefont {E.~J.}\ \bibnamefont {Miller}}, \bibinfo
  {author} {\bibfnamefont {B.}~\bibnamefont {Bozkurt}}, \bibinfo {author}
  {\bibfnamefont {K.}~\bibnamefont {Ya{\u{g}}murlu}}, \bibinfo {author}
  {\bibfnamefont {E.~C.}\ \bibnamefont {Woolf}}, \bibinfo {author}
  {\bibfnamefont {A.~C.}\ \bibnamefont {Scheck}}, \bibinfo {author}
  {\bibfnamefont {J.~M.}\ \bibnamefont {Eschbacher}}, \bibinfo {author}
  {\bibfnamefont {P.}~\bibnamefont {Nakaji}}, \ and\ \bibinfo {author}
  {\bibfnamefont {M.~C.}\ \bibnamefont {Preul}},\ }\href {\doibase
  10.2147/CMAR.S165980} {\bibfield  {journal} {\bibinfo  {journal} {Cancer
  management and research}\ }\textbf {\bibinfo {volume} {10}},\ \bibinfo
  {pages} {3109} (\bibinfo {year} {2018}{\natexlab{b}})}\BibitemShut {NoStop}%
\bibitem [{\citenamefont {Lombardini}\ \emph {et~al.}(2018)\citenamefont
  {Lombardini}, \citenamefont {Mytskaniuk}, \citenamefont {Sivankutty},
  \citenamefont {Andresen}, \citenamefont {Chen}, \citenamefont {Wenger},
  \citenamefont {Fabert}, \citenamefont {Joly}, \citenamefont {Louradour},
  \citenamefont {Kudlinski} \emph {et~al.}}]{lombardini2018high}%
  \BibitemOpen
  \bibfield  {author} {\bibinfo {author} {\bibfnamefont {A.}~\bibnamefont
  {Lombardini}}, \bibinfo {author} {\bibfnamefont {V.}~\bibnamefont
  {Mytskaniuk}}, \bibinfo {author} {\bibfnamefont {S.}~\bibnamefont
  {Sivankutty}}, \bibinfo {author} {\bibfnamefont {E.~R.}\ \bibnamefont
  {Andresen}}, \bibinfo {author} {\bibfnamefont {X.}~\bibnamefont {Chen}},
  \bibinfo {author} {\bibfnamefont {J.}~\bibnamefont {Wenger}}, \bibinfo
  {author} {\bibfnamefont {M.}~\bibnamefont {Fabert}}, \bibinfo {author}
  {\bibfnamefont {N.}~\bibnamefont {Joly}}, \bibinfo {author} {\bibfnamefont
  {F.}~\bibnamefont {Louradour}}, \bibinfo {author} {\bibfnamefont
  {A.}~\bibnamefont {Kudlinski}},  \emph {et~al.},\ }\href {\doibase
  10.1038/s41377-018-0003-3} {\bibfield  {journal} {\bibinfo  {journal} {Light:
  Science \& Applications}\ }\textbf {\bibinfo {volume} {7}},\ \bibinfo {pages}
  {10} (\bibinfo {year} {2018})}\BibitemShut {NoStop}%
\bibitem [{\citenamefont {Stummer}\ \emph {et~al.}(2006)\citenamefont
  {Stummer}, \citenamefont {Pichlmeier}, \citenamefont {Meinel}, \citenamefont
  {Wiestler}, \citenamefont {Zanella}, \citenamefont {Reulen}, \citenamefont
  {Group} \emph {et~al.}}]{stummer2006fluorescence}%
  \BibitemOpen
  \bibfield  {author} {\bibinfo {author} {\bibfnamefont {W.}~\bibnamefont
  {Stummer}}, \bibinfo {author} {\bibfnamefont {U.}~\bibnamefont {Pichlmeier}},
  \bibinfo {author} {\bibfnamefont {T.}~\bibnamefont {Meinel}}, \bibinfo
  {author} {\bibfnamefont {O.~D.}\ \bibnamefont {Wiestler}}, \bibinfo {author}
  {\bibfnamefont {F.}~\bibnamefont {Zanella}}, \bibinfo {author} {\bibfnamefont
  {H.-J.}\ \bibnamefont {Reulen}}, \bibinfo {author} {\bibfnamefont {A.-G.~S.}\
  \bibnamefont {Group}},  \emph {et~al.},\ }\href@noop {} {\bibfield  {journal}
  {\bibinfo  {journal} {The lancet oncology}\ }\textbf {\bibinfo {volume}
  {7}},\ \bibinfo {pages} {392} (\bibinfo {year} {2006})}\BibitemShut {NoStop}%
\bibitem [{\citenamefont {Lippok}\ \emph {et~al.}(2019)\citenamefont {Lippok},
  \citenamefont {Siddiqui}, \citenamefont {Vakoc},\ and\ \citenamefont
  {Bouma}}]{Lippok2018}%
  \BibitemOpen
  \bibfield  {author} {\bibinfo {author} {\bibfnamefont {N.}~\bibnamefont
  {Lippok}}, \bibinfo {author} {\bibfnamefont {M.}~\bibnamefont {Siddiqui}},
  \bibinfo {author} {\bibfnamefont {B.~J.}\ \bibnamefont {Vakoc}}, \ and\
  \bibinfo {author} {\bibfnamefont {B.~E.}\ \bibnamefont {Bouma}},\ }\href
  {\doibase 10.1103/PhysRevApplied.11.014018} {\bibfield  {journal} {\bibinfo
  {journal} {Phys. Rev. Applied}\ }\textbf {\bibinfo {volume} {11}},\ \bibinfo
  {pages} {014018} (\bibinfo {year} {2019})}\BibitemShut {NoStop}%
\bibitem [{\citenamefont {Bachmann}\ \emph {et~al.}(2006)\citenamefont
  {Bachmann}, \citenamefont {Leitgeb},\ and\ \citenamefont
  {Lasser}}]{Bachmann:06}%
  \BibitemOpen
  \bibfield  {author} {\bibinfo {author} {\bibfnamefont {A.~H.}\ \bibnamefont
  {Bachmann}}, \bibinfo {author} {\bibfnamefont {R.~A.}\ \bibnamefont
  {Leitgeb}}, \ and\ \bibinfo {author} {\bibfnamefont {T.}~\bibnamefont
  {Lasser}},\ }\href {\doibase 10.1364/OE.14.001487} {\bibfield  {journal}
  {\bibinfo  {journal} {Opt. Express}\ }\textbf {\bibinfo {volume} {14}},\
  \bibinfo {pages} {1487} (\bibinfo {year} {2006})}\BibitemShut {NoStop}%
\bibitem [{\citenamefont {Karpov}\ \emph {et~al.}(2018)\citenamefont {Karpov},
  \citenamefont {Pfeiffer}, \citenamefont {Liu}, \citenamefont {Lukashchuk},\
  and\ \citenamefont {Kippenberg}}]{karpov_photonic_2018}%
  \BibitemOpen
  \bibfield  {author} {\bibinfo {author} {\bibfnamefont {M.}~\bibnamefont
  {Karpov}}, \bibinfo {author} {\bibfnamefont {M.~H.~P.}\ \bibnamefont
  {Pfeiffer}}, \bibinfo {author} {\bibfnamefont {J.}~\bibnamefont {Liu}},
  \bibinfo {author} {\bibfnamefont {A.}~\bibnamefont {Lukashchuk}}, \ and\
  \bibinfo {author} {\bibfnamefont {T.~J.}\ \bibnamefont {Kippenberg}},\ }\href
  {\doibase 10.1038/s41467-018-03471-x} {\bibfield  {journal} {\bibinfo
  {journal} {Nature Communications}\ }\textbf {\bibinfo {volume} {9}} (\bibinfo
  {year} {2018}),\ 10.1038/s41467-018-03471-x}\BibitemShut {NoStop}%
\bibitem [{\citenamefont {Lee}\ \emph {et~al.}(2017)\citenamefont {Lee},
  \citenamefont {Oh}, \citenamefont {Yang}, \citenamefont {Shen}, \citenamefont
  {Wang}, \citenamefont {Yang}, \citenamefont {Lai}, \citenamefont {Yi},
  \citenamefont {Li},\ and\ \citenamefont {Vahala}}]{lee2017towards}%
  \BibitemOpen
  \bibfield  {author} {\bibinfo {author} {\bibfnamefont {S.~H.}\ \bibnamefont
  {Lee}}, \bibinfo {author} {\bibfnamefont {D.~Y.}\ \bibnamefont {Oh}},
  \bibinfo {author} {\bibfnamefont {Q.-F.}\ \bibnamefont {Yang}}, \bibinfo
  {author} {\bibfnamefont {B.}~\bibnamefont {Shen}}, \bibinfo {author}
  {\bibfnamefont {H.}~\bibnamefont {Wang}}, \bibinfo {author} {\bibfnamefont
  {K.~Y.}\ \bibnamefont {Yang}}, \bibinfo {author} {\bibfnamefont {Y.-H.}\
  \bibnamefont {Lai}}, \bibinfo {author} {\bibfnamefont {X.}~\bibnamefont
  {Yi}}, \bibinfo {author} {\bibfnamefont {X.}~\bibnamefont {Li}}, \ and\
  \bibinfo {author} {\bibfnamefont {K.}~\bibnamefont {Vahala}},\ }\href@noop {}
  {\bibfield  {journal} {\bibinfo  {journal} {Nature communications}\ }\textbf
  {\bibinfo {volume} {8}},\ \bibinfo {pages} {1295} (\bibinfo {year}
  {2017})}\BibitemShut {NoStop}%
\bibitem [{\citenamefont {Pfeiffer}\ \emph
  {et~al.}(2018{\natexlab{a}})\citenamefont {Pfeiffer}, \citenamefont
  {Herkommer}, \citenamefont {Liu}, \citenamefont {Morais}, \citenamefont
  {Zervas}, \citenamefont {Geiselmann},\ and\ \citenamefont
  {Kippenberg}}]{Pfeiffer:18a}%
  \BibitemOpen
  \bibfield  {author} {\bibinfo {author} {\bibfnamefont {M.~H.~P.}\
  \bibnamefont {Pfeiffer}}, \bibinfo {author} {\bibfnamefont {C.}~\bibnamefont
  {Herkommer}}, \bibinfo {author} {\bibfnamefont {J.}~\bibnamefont {Liu}},
  \bibinfo {author} {\bibfnamefont {T.}~\bibnamefont {Morais}}, \bibinfo
  {author} {\bibfnamefont {M.}~\bibnamefont {Zervas}}, \bibinfo {author}
  {\bibfnamefont {M.}~\bibnamefont {Geiselmann}}, \ and\ \bibinfo {author}
  {\bibfnamefont {T.~J.}\ \bibnamefont {Kippenberg}},\ }\bibfield  {booktitle}
  {\emph {\bibinfo {booktitle} {IEEE Journal of Selected Topics in Quantum
  Electronics}},\ }\href {\doibase 10.1109/JSTQE.2018.2808258} {\bibfield
  {journal} {\bibinfo  {journal} {IEEE Journal of Selected Topics in Quantum
  Electronics}\ }\textbf {\bibinfo {volume} {24}},\ \bibinfo {pages} {1}
  (\bibinfo {year} {2018}{\natexlab{a}})}\BibitemShut {NoStop}%
\bibitem [{\citenamefont {Liu}\ \emph {et~al.}(2018{\natexlab{b}})\citenamefont
  {Liu}, \citenamefont {Raja}, \citenamefont {Pfeiffer}, \citenamefont
  {Herkommer}, \citenamefont {Guo}, \citenamefont {Zervas}, \citenamefont
  {Geiselmann},\ and\ \citenamefont {Kippenberg}}]{Liu:18}%
  \BibitemOpen
  \bibfield  {author} {\bibinfo {author} {\bibfnamefont {J.}~\bibnamefont
  {Liu}}, \bibinfo {author} {\bibfnamefont {A.~S.}\ \bibnamefont {Raja}},
  \bibinfo {author} {\bibfnamefont {M.~H.~P.}\ \bibnamefont {Pfeiffer}},
  \bibinfo {author} {\bibfnamefont {C.}~\bibnamefont {Herkommer}}, \bibinfo
  {author} {\bibfnamefont {H.}~\bibnamefont {Guo}}, \bibinfo {author}
  {\bibfnamefont {M.}~\bibnamefont {Zervas}}, \bibinfo {author} {\bibfnamefont
  {M.}~\bibnamefont {Geiselmann}}, \ and\ \bibinfo {author} {\bibfnamefont
  {T.~J.}\ \bibnamefont {Kippenberg}},\ }\href {\doibase 10.1364/OL.43.003200}
  {\bibfield  {journal} {\bibinfo  {journal} {Opt. Lett.}\ }\textbf {\bibinfo
  {volume} {43}},\ \bibinfo {pages} {3200} (\bibinfo {year}
  {2018}{\natexlab{b}})}\BibitemShut {NoStop}%
\bibitem [{\citenamefont {Pfeiffer}\ \emph
  {et~al.}(2018{\natexlab{b}})\citenamefont {Pfeiffer}, \citenamefont {Liu},
  \citenamefont {Raja}, \citenamefont {Morais}, \citenamefont {Ghadiani},\ and\
  \citenamefont {Kippenberg}}]{pfeiffer_ultra-smooth_2018}%
  \BibitemOpen
  \bibfield  {author} {\bibinfo {author} {\bibfnamefont {M.~H.~P.}\
  \bibnamefont {Pfeiffer}}, \bibinfo {author} {\bibfnamefont {J.}~\bibnamefont
  {Liu}}, \bibinfo {author} {\bibfnamefont {A.~S.}\ \bibnamefont {Raja}},
  \bibinfo {author} {\bibfnamefont {T.}~\bibnamefont {Morais}}, \bibinfo
  {author} {\bibfnamefont {B.}~\bibnamefont {Ghadiani}}, \ and\ \bibinfo
  {author} {\bibfnamefont {T.~J.}\ \bibnamefont {Kippenberg}},\ }\href
  {\doibase 10.1364/OPTICA.5.000884} {\bibfield  {journal} {\bibinfo  {journal}
  {Optica, OPTICA}\ }\textbf {\bibinfo {volume} {5}},\ \bibinfo {pages} {884}
  (\bibinfo {year} {2018}{\natexlab{b}})}\BibitemShut {NoStop}%
\bibitem [{\citenamefont {Marchand}\ \emph {et~al.}(2018)\citenamefont
  {Marchand}, \citenamefont {Szlag}, \citenamefont {Extermann}, \citenamefont
  {Bouwens}, \citenamefont {Nguyen}, \citenamefont {Rudin},\ and\ \citenamefont
  {Lasser}}]{Marchand2018}%
  \BibitemOpen
  \bibfield  {author} {\bibinfo {author} {\bibfnamefont {P.~J.}\ \bibnamefont
  {Marchand}}, \bibinfo {author} {\bibfnamefont {D.}~\bibnamefont {Szlag}},
  \bibinfo {author} {\bibfnamefont {J.}~\bibnamefont {Extermann}}, \bibinfo
  {author} {\bibfnamefont {A.}~\bibnamefont {Bouwens}}, \bibinfo {author}
  {\bibfnamefont {D.}~\bibnamefont {Nguyen}}, \bibinfo {author} {\bibfnamefont
  {M.}~\bibnamefont {Rudin}}, \ and\ \bibinfo {author} {\bibfnamefont
  {T.}~\bibnamefont {Lasser}},\ }\href {\doibase 10.1364/OL.43.001782}
  {\bibfield  {journal} {\bibinfo  {journal} {Optics Letters}\ }\textbf
  {\bibinfo {volume} {43}},\ \bibinfo {pages} {1782} (\bibinfo {year}
  {2018})}\BibitemShut {NoStop}%
\bibitem [{\citenamefont {Schindelin}\ \emph {et~al.}(2012)\citenamefont
  {Schindelin}, \citenamefont {Arganda-Carreras}, \citenamefont {Frise},
  \citenamefont {Kaynig}, \citenamefont {Longair}, \citenamefont {Pietzsch},
  \citenamefont {Preibisch}, \citenamefont {Rueden}, \citenamefont {Saalfeld},
  \citenamefont {Schmid}, \citenamefont {Tinevez}, \citenamefont {White},
  \citenamefont {Hartenstein}, \citenamefont {Eliceiri}, \citenamefont
  {Tomancak},\ and\ \citenamefont {Cardona}}]{Schindelin2012}%
  \BibitemOpen
  \bibfield  {author} {\bibinfo {author} {\bibfnamefont {J.}~\bibnamefont
  {Schindelin}}, \bibinfo {author} {\bibfnamefont {I.}~\bibnamefont
  {Arganda-Carreras}}, \bibinfo {author} {\bibfnamefont {E.}~\bibnamefont
  {Frise}}, \bibinfo {author} {\bibfnamefont {V.}~\bibnamefont {Kaynig}},
  \bibinfo {author} {\bibfnamefont {M.}~\bibnamefont {Longair}}, \bibinfo
  {author} {\bibfnamefont {T.}~\bibnamefont {Pietzsch}}, \bibinfo {author}
  {\bibfnamefont {S.}~\bibnamefont {Preibisch}}, \bibinfo {author}
  {\bibfnamefont {C.}~\bibnamefont {Rueden}}, \bibinfo {author} {\bibfnamefont
  {S.}~\bibnamefont {Saalfeld}}, \bibinfo {author} {\bibfnamefont
  {B.}~\bibnamefont {Schmid}}, \bibinfo {author} {\bibfnamefont {J.-Y.}\
  \bibnamefont {Tinevez}}, \bibinfo {author} {\bibfnamefont {D.~J.}\
  \bibnamefont {White}}, \bibinfo {author} {\bibfnamefont {V.}~\bibnamefont
  {Hartenstein}}, \bibinfo {author} {\bibfnamefont {K.}~\bibnamefont
  {Eliceiri}}, \bibinfo {author} {\bibfnamefont {P.}~\bibnamefont {Tomancak}},
  \ and\ \bibinfo {author} {\bibfnamefont {A.}~\bibnamefont {Cardona}},\ }\href
  {https://doi.org/10.1038/nmeth.2019 http://10.0.4.14/nmeth.2019
  https://www.nature.com/articles/nmeth.2019{\#}supplementary-information}
  {\bibfield  {journal} {\bibinfo  {journal} {Nature Methods}\ }\textbf
  {\bibinfo {volume} {9}},\ \bibinfo {pages} {676} (\bibinfo {year}
  {2012})}\BibitemShut {NoStop}%
\bibitem [{\citenamefont {Marchand}\ \emph {et~al.}(2017)\citenamefont
  {Marchand}, \citenamefont {Bouwens}, \citenamefont {Szlag}, \citenamefont
  {Nguyen}, \citenamefont {Descloux}, \citenamefont {Sison}, \citenamefont
  {Coquoz}, \citenamefont {Extermann},\ and\ \citenamefont
  {Lasser}}]{Marchand2017a}%
  \BibitemOpen
  \bibfield  {author} {\bibinfo {author} {\bibfnamefont {P.~J.}\ \bibnamefont
  {Marchand}}, \bibinfo {author} {\bibfnamefont {A.}~\bibnamefont {Bouwens}},
  \bibinfo {author} {\bibfnamefont {D.}~\bibnamefont {Szlag}}, \bibinfo
  {author} {\bibfnamefont {D.}~\bibnamefont {Nguyen}}, \bibinfo {author}
  {\bibfnamefont {A.}~\bibnamefont {Descloux}}, \bibinfo {author}
  {\bibfnamefont {M.}~\bibnamefont {Sison}}, \bibinfo {author} {\bibfnamefont
  {S.}~\bibnamefont {Coquoz}}, \bibinfo {author} {\bibfnamefont
  {J.}~\bibnamefont {Extermann}}, \ and\ \bibinfo {author} {\bibfnamefont
  {T.}~\bibnamefont {Lasser}},\ }\href {\doibase 10.1364/BOE.8.003343}
  {\bibfield  {journal} {\bibinfo  {journal} {Biomed. Opt. Express}\ }\textbf
  {\bibinfo {volume} {8}},\ \bibinfo {pages} {3343} (\bibinfo {year}
  {2017})}\BibitemShut {NoStop}%
\bibitem [{\citenamefont {Nguyen}\ \emph {et~al.}(2017)\citenamefont {Nguyen},
  \citenamefont {Marchand}, \citenamefont {Planchette}, \citenamefont
  {Nilsson}, \citenamefont {Sison}, \citenamefont {Extermann}, \citenamefont
  {Lopez}, \citenamefont {Sylwestrzak}, \citenamefont {Sordet-Dessimoz},
  \citenamefont {Schmidt-Christensen}, \citenamefont {Holmberg}, \citenamefont
  {{Van De Ville}},\ and\ \citenamefont {Lasser}}]{Nguyen2017}%
  \BibitemOpen
  \bibfield  {author} {\bibinfo {author} {\bibfnamefont {D.}~\bibnamefont
  {Nguyen}}, \bibinfo {author} {\bibfnamefont {P.~J.}\ \bibnamefont
  {Marchand}}, \bibinfo {author} {\bibfnamefont {A.~L.}\ \bibnamefont
  {Planchette}}, \bibinfo {author} {\bibfnamefont {J.}~\bibnamefont {Nilsson}},
  \bibinfo {author} {\bibfnamefont {M.}~\bibnamefont {Sison}}, \bibinfo
  {author} {\bibfnamefont {J.}~\bibnamefont {Extermann}}, \bibinfo {author}
  {\bibfnamefont {A.}~\bibnamefont {Lopez}}, \bibinfo {author} {\bibfnamefont
  {M.}~\bibnamefont {Sylwestrzak}}, \bibinfo {author} {\bibfnamefont
  {J.}~\bibnamefont {Sordet-Dessimoz}}, \bibinfo {author} {\bibfnamefont
  {A.}~\bibnamefont {Schmidt-Christensen}}, \bibinfo {author} {\bibfnamefont
  {D.}~\bibnamefont {Holmberg}}, \bibinfo {author} {\bibfnamefont
  {D.}~\bibnamefont {{Van De Ville}}}, \ and\ \bibinfo {author} {\bibfnamefont
  {T.}~\bibnamefont {Lasser}},\ }\href {\doibase 10.1364/BOE.8.005637}
  {\bibfield  {journal} {\bibinfo  {journal} {Biomed. Opt. Express}\ }\textbf
  {\bibinfo {volume} {8}},\ \bibinfo {pages} {5637} (\bibinfo {year}
  {2017})}\BibitemShut {NoStop}%
\end{thebibliography}%

\section*{Methods}
Here we describe the experimental realization of Kerr comb based SD-OCT. Figure \ref{fig_combSource} shows the experimental setting consisting of two distinct setups located in buildings spaced by about $\sim 700\,\mathrm{m}$. A fiber link connects the setup for DKS generation and the SD-OCT setup between the two laboratories. \\

\textbf{Micro-resonator fabrication.} 
The samples employed are $1\,\mathrm{THz}$ FSR micro-resonators formed by \ce{Si3N4} waveguides. Figure \ref{fig_octPrinciple} e) shows the micro-resonator used in this work. These resonators were fabricated using the photonic Damascene process which avoids common processing challenges of thick \ce{Si3N4} films \cite{Pfeiffer2016, Pfeiffer:18a} and has recently allowed for micro-resonator Q factors exceeding 10 million \cite{Liu:18a}. The continuous wave pumping light is coupled into the \ce{Si3N4} chips via a double inverse taper \cite{Liu:18}. For the 1 THz DKS comb, the cross section is $1.45 \times 0.78 \,\mathrm{\mu m^2}$ while for the chaotic comb, it is $1.425 \times 0.73\,\mathrm{\mu m^2}$. The bus waveguides (design width $0.55\,\mathrm{\mu m}$ for DKS and $0.525\,\mathrm{\mu m}$ for the chaotic comb) couple the light into the ring resonators ($22.71\,\mathrm{\mu m}$ radius) are mode matched to excite the fundamental $\mathrm{TM_{00}}$ mode. The resonance linewidth is below $100 \, \mathrm{MHz}$ as has been measured in the recent publication \cite{pfeiffer_ultra-smooth_2018}. The waveguide cross-section of the 100 and 200~GHz FSR DKS microresonators is $1.52 \times 0.82 \,\mathrm{\mu m^2}$. The simulated dispersion profiles, including the modal dispersion ($D_\mathrm{2}$) and the modal deviation from the resonance frequency of the nearest mode ($D_\mathrm{int}$) can be found in figure \ref{fig_dispersion}. \\

\textbf{Kerr comb generation.}
The DKS light source is pumped by a $1300\,\mathrm{nm}$ external cavity diode laser, which is amplified up to $\sim 650\,\mathrm{mW}$ power using a semiconductor optical amplifier (SOA). The amplified light is coupled to the silicon nitride micro-resonator chip via lensed fibers. The pump polarization can be adjusted via a paddle controller and both the power before and after the chip are monitored via power meters (PM). We estimate a soliton excitation power in the bus waveguide of $\sim290\,\mathrm{mW}$. An arbitrary function generator (AFG) provides the voltage ramp signal driving the laser frequency tuning.  A standard voltage ramp tuning method \cite{Guo2016}: through the voltage-ramp tuning, a multi-soliton state is excited which is then converted into a single soliton through backward tuning. A tunable fiber Bragg grating (FBG) is used to attenuate the residual pump light while the back-reflected pump is detected through a fast photodiode and shown on an oscilloscope (OSC) to monitor the laser tuning. The generated DKS spectrum is free of avoided mode crossings causing strong local power deviations typically originating from the multimodal nature of the waveguide. \\

\textbf{OCT imaging.} 
The generated DKS comb is then sent to a custom-built OCT setup through a $\sim 700\,\mathrm{m}$ long optical fiber link (to connect the source from one laboratory to the OCT setup in another laboratory). The SD-OCT setup was designed for a commercial SLD with a central wavelength $\lambda_0 = 1310\,\mathrm{nm}$ and bandwidth $\delta\lambda = 150\,\mathrm{nm}$ (LS2000C, Thorlabs) and its detection is based on a highly sensitive spectrometer, as described previously \cite{Marchand2018}. Both the source and detection are connected via a broadband fiber beam splitter (TW1300R5A2, Thorlabs) with a dispersion compensated reference arm and a sample arm comprising a galvo mirror scan unit (6210H, Cambridge Technologies), a high NA objective (LUMPLFLN-40XW, Olympus) and imaging optics. The scanner control and data readout are performed by a connected computer with a high-speed input. The output optical power of the SLD source is $\sim 9.15 \, \mathrm{mW}$ while the DKS comb is $<2.3 \,\mathrm{mW}$. All images were acquired at an A-scan rate of $46\,\mathrm{kHz}$. The post-processing steps, including k-space resampling and Fourier transformations were performed using a custom software implemented in MATLAB (Mathworks). The axial resolution of the SLD and DKS systems were characterized by placing a mirror in the front focal plane of the objective and were measured as $\sim 6\,\mathrm{\mu m}$ and $\sim 10\,\mathrm{\mu m}$ in air respectively.\\

\textbf{Image processing.}
The images presented in Fig. \ref{fig_imagingResults} were obtained after Fourier transform of the spectral interferograms recorded by the spectrometer. Prior to visualization, the dynamic range of the data was reduced using first a logarithmic operation (10$\times$log$_{10}$()) and a clipping operation (same operations for both DKS and SLD images). The data was then spatially smoothed using a median filter in MATLAB \cite{Schindelin2012}, planes at different depths were selected. The clipping limits were obtained by taking the 0.01\% and 99.9\% intensity values of the imaged planes, after median filtering.\\ 
Background subtraction was performed, prior to Fourier transforming, by averaging each B-scan into a single background vector, which was then subtracted to the entire B-scan. This step was repeated for each B-scan of the volume. This step was performed for both SLD and DKS data. The DKS data was processed in two separate ways, as shown in Fig. \ref{fig_imagingResults}, either by considering the entire interferogram or by selecting only the comb peaks. In the first processing method, the entire interferogram was considered, including non-illuminated pixels. The obtained A-scan for each position is of the same length as the spectral interferogram (Fig. \ref{fig_imagingResults} e). In the second method, the comb tones positions on the interferogram were first identified by computing the local maximal value around the tone. Using their positions, shorter interferogram were obtained by eliminating all other pixels (non comb tone pixels). The resulting A-scan were therefore significantly shorter than those obtained in the first method, and include solely one ambiguity range (Fig. \ref{fig_imagingResults} c-d). The dynamic range of the planes for both methods was 19 and 23 dB for the first and second processing pipelines respectively.\\

\textbf{Brain tissue preparation.}
All animal procedures were carried out according to Swiss regulations under the approval of the veterinary authority of the canton of Vaud (protocols VD3056 and VD3058), are in-line with the 3Rs and follow the ARRIVE guidelines. After transcardiac perfusion, the brains of B6SJL/f1 mice were extracted, placed into 4\% PFA overnight and then placed in a solution of 30\% glucose. The brains were finally cut into slices of $\sim$ $50\,\mathrm{\mu m}$ using a microtome and placed on a glass coverslide. These samples had been prepared for previous studies \cite{Marchand2017a, Nguyen2017}, no new samples were prepared for this manuscript.\\

\newpage
\clearpage

\section*{Supporting information}
\begin{figure*}[!h]
\includegraphics[width = 0.9\textwidth]{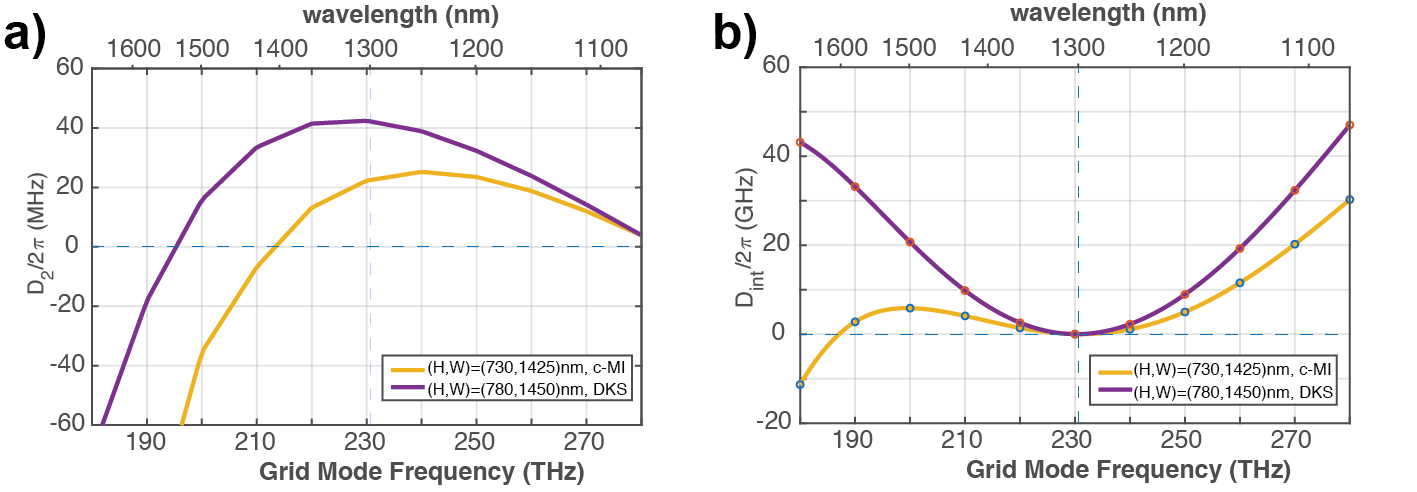}
\caption{\textbf{Dispersion simulations of the micro-resonators} a) GVD parameter ($\mathrm{D_2/2\pi}$) and b) integrated dispersion ($\mathrm{D_{int}/2\pi}$) of the micro-resonators were simulated using COMSOL multiphysics\textsuperscript{\textregistered} simulation package with the given dimension parameters detailed in the Methods section}
\label{fig_dispersion}
\end{figure*} 
The dispersion profiles, including the GVD parameter ($\mathrm{D_2/2\pi}$) and the integrated dispersion ($\mathrm{D_{int}/2\pi}$) (Fig. \ref{fig_dispersion}) of the micro-resonators were simulated using COMSOL multiphysics\textsuperscript{\textregistered} with the 2D axial symmetric model. The cross-section dimension is $1.45 \times 0.78\,\mathrm{\mu m^2}$ for the DKS comb and $1.425 \times 0.73\,\mathrm{\mu m^2}$ for the chaotic comb. Both the radius of the resonators were $22.71\,\mathrm{\mu m}$. The $\mathrm{TM_{00}}$ mode in the waveguides were calculated. \\ 

\begin{figure*}[!h]
\includegraphics[width = 0.9\textwidth]{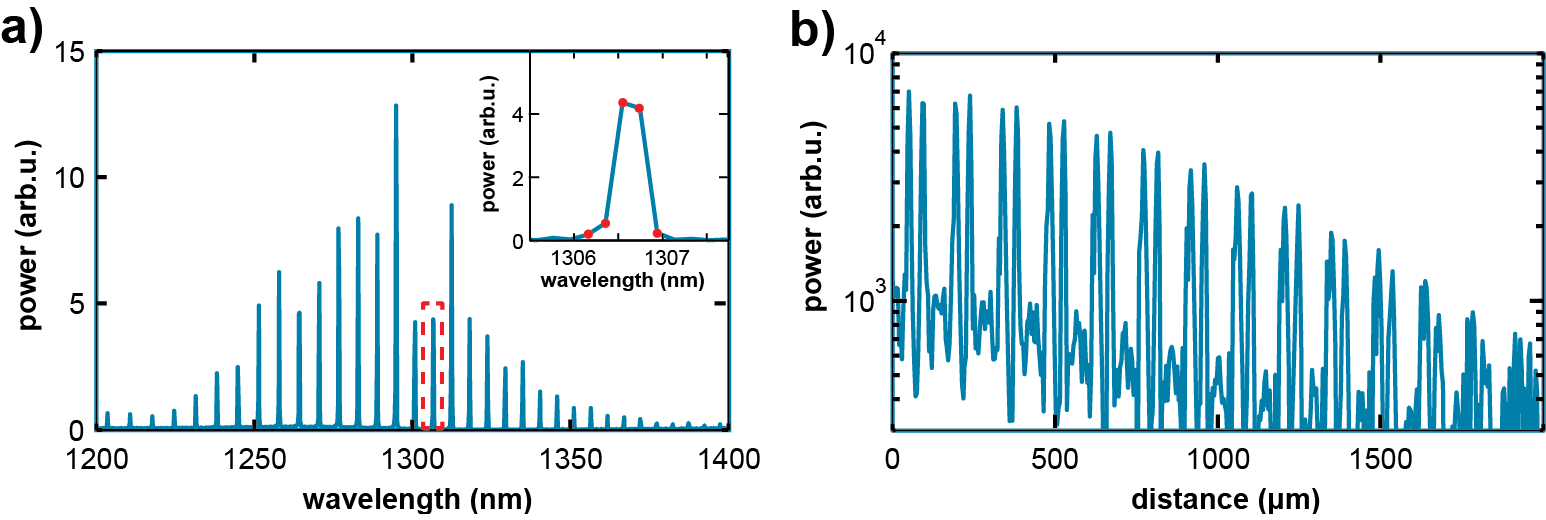}
\caption{\textbf{Characterization of imaging performance} a) DKS comb spectrum as acquired by the spectrometer's line sensor. Inset showing a zoomed-in view of a single tooth of the comb being sampled by adjacent CCD pixels. b) Tomogram of the a mirror placed under the objective obtained with the DKS source, obtained after re-sampling and Fourier transformation of the spectral interferogram.}
\label{fig_mirror}
\end{figure*} 




To characterize the imaging performance of the DKS as a light source for OCT imaging, we used a highly reflective mirror substrate as a sample. Figure \ref{fig_mirror} a) shows the DKS signal as recorded by the spectrometer's image sensor without the mirror (i.e. signal from the reference arm). Interestingly, although the line width is significantly shorter than the camera's spectral sampling, the comb line can be sampled on  adjacent pixels as can be seen in the inset. From the Gaussian-like shape of the recorded peak, we believe this effect to be likely caused by the diffraction limited size of the spot on the camera. Other potential causes include electronic cross-talk between the pixels and a sub-optimal matching between comb teeth spacing and the detector pixel pitch. The apparent divergence between this observation and the conceptual illustration presented in Fig. \ref{fig_octPrinciple} a) does not however impact the circular ranging abilities of the source, as the coherence length of a single comb tone still exceeds the imaging range of the spectrometer. After placing the mirror, we obtain the tomogram shown in logarithmic scale in Figure \ref{fig_mirror} b) by Fourier transforming the signal obtained from the spectrometer. A resolution of $\sim 10\,\mathrm{\mu m}$ and an ambiguity range of $\sim 71\,\mathrm{\mu m}$ are derived in the air. The signal attenuation over depth observed, also termed roll-off, is caused primarily by the spectrometer's finite spectral sampling despite the fine width of the comb's lines.

\end{document}